\documentclass[journal]{IEEEtran} 

\usepackage[utf8]{inputenc}
\usepackage{amsmath,amssymb,amsfonts}
\usepackage[numbers,sort&compress]{natbib}
\usepackage{bm}
\usepackage{color}
\usepackage{glossaries}
\usepackage{graphicx}
\usepackage[linesnumbered]{algorithm2e}
\usepackage{algpseudocode}
\usepackage[caption = false]{subfig}
\usepackage{url}
\newacronym{BS}{BS}{base station}
\newacronym{CDF}{CDF}{cumulative distribution function}
\newacronym{FAMA}{FAMA}{fluid antenna multiple access}
\newacronym{FAS}{FAS}{fluid antenna system}
\newacronym{OP}{OP}{outage probability}
\newacronym{PDF}{PDF}{probability density function}
\newacronym{SIR}{SIR}{signal-to-interference ratio}
\newacronym{SNR}{SNR}{signal-to-noise ratio}
\renewcommand{\exp}[1]{\text{exp}\left(#1\right)} 
\newcommand{\besseli}[1]{I_0\left(#1\right)} 
\newcommand{\dif}{\,\text{d}}					  
\renewcommand{\vec}[1]{\mathbf{\lowercase{#1}}}	   
\newcommand{\mat}[1]{\mathbf{\uppercase{#1}}}	   
\newcommand{\rev}[1]{#1}

\newtheorem{lemma}{Lemma}
\newtheorem{corollary}{Corollary}
\newtheorem{remark}{Remark}

\title{A New Spatial Block-Correlation Model for Fluid Antenna Systems}
\author{Pablo Ram\'irez-Espinosa, David Morales-Jimenez, \emph{Senior Member, IEEE,} and Kai-Kit Wong, \emph{Fellow, IEEE}  
\thanks{This work is supported in part by the State Research Agency (AEI) of Spain under grant
PID2020-118139RB-I0/AEI/10.13039/501100011033. The work of P. Ram\'irez-Espinosa is supported by a ``Mar\'ia Zambrano" Fellowship funded by the European Union – Next Generation EU via the Ministry of Universities of the Spanish Government. The work of D. Morales-Jimenez is supported by AEI and the European Social Fund under grant RYC2020- 030536-I. The work of K. K. Wong is supported by the Engineering and Physical Sciences Research Council (EPSRC) under Grant EP/W026813/1.}
\thanks{P. Ramírez-Espinosa is with Communications Engineering Department, University of M\'alaga, M\'alaga 29071, Spain (e-mail: pre@ic.uma.es).}
\thanks{D. Morales-Jimenez is with Dept. Signal Theory, Networking and Communications, Universidad de Granada, Granada 18071, Spain (e-mail: dmorales@ugr.es).}
\thanks{K.-K. Wong is with the Department of Electronic and Electrical Engineering, University College London, London WC1E 6BT, U.K. (e-mail: kai-kit.wong@ucl.ac.uk). He is also affiliated with Yonsei Frontier Laboratory, Yonsei University, Seoul, 03722, Korea.}
\thanks{This work has been submitted to the IEEE for publication. Copyright may be transferred without notice, after which this version may no longer be accesible.}
}

\begin{document}
\maketitle

\begin{abstract}
    Powered by position-flexible antennas, the emerging fluid antenna system (FAS) technology \rev{is postulated} as a key enabler for massive connectivity in 6G networks. The free movement of antenna elements enables the opportunistic minimization of interference, allowing several users to share the same radio channel without the need of precoding. However, the true potential of FAS is still unknown due to the extremely high spatial correlation of the wireless channel between very close-by antenna positions. To unveil the multiplexing capabilities of FAS, proper (simple yet accurate) modeling of the spatial correlation is prominently needed. Realistic classical models such as Jakes's are prohibitively complex, rendering intractable analyses, while state-of-the-art approximations often are too simplistic and poorly accurate. Aiming to fill this gap, we here propose a general framework to approximate spatial correlation by block-diagonal matrices, motivated by the well-known block fading assumption and by statistical results on large correlation matrices. The proposed block-correlation model makes the performance analysis possible, and tightly approximates the results obtained with realistic models (Jakes's and Clarke's). Our framework is leveraged to analyze fluid antenna multiple access (FAMA) systems,
    evaluating their performance for both one- and two-dimensional fluid antennas.  
\end{abstract}

%------------------------------------------------------------------------------------------------------------------
% Introduction
%------------------------------------------------------------------------------------------------------------------
\section{Introduction}

\Gls{FAS} is an emerging and very promising technology, focused on dynamic antenna architectures, which is recently attracting substantial interest as a key enabler for massive multiple access in upcoming 6G systems \cite{Wong2022_BruceLee}. Conceptually, \gls{FAS} encapsulates any system in which the antenna elements can be moved within a predefined aperture, either virtually or physically. This includes all forms of position-flexible antennas such as on-off switching pixels \cite{Besoli2011, Cetiner2004}, liquid-based radiating structures \cite{Huang2021}, or movable antenna systems as recently proposed in \rev{\cite{Zhu2023, Zhu2023Magazine}}; see these works and the references therein for more details on physical implementations. 

The key idea behind \gls{FAS} is similar to that in classical diversity systems with fixed antennas, i.e., exploiting the spatial diversity to minimize interference or maximize the received signal strength. However, in contrast to conventional fixed arrays, the \emph{antennas} in \gls{FAS} (more rigorously, \emph{radiating elements})
%since not all physical implementations involve conventional antennas
can be freely moved/switched to any desired position within the given aperture, thus allowing to fully exploit spatial diversity. Indeed, the additional degrees of freedom enable the space (continuum) to be ``sampled" at any arbitrary point, albeit in practice only a finite---arbitrarily dense---mesh of points may be available. Remark that, while \gls{FAS} was originally envisioned as one-dimensional (1D), i.e., antenna elements moving along a line, the same concept can be extended to 2D structures. In fact, preliminary results in \cite{Kiat2023} suggest that the degrees of freedom brought by \gls{FAS} may be ramped up in planar apertures, where the convenient point (e.g., lowest interference) is sought within a 2D surface, rather than along a line.

Based on the dynamic capabilities of \gls{FAS} (i.e., how fast the radiating elements can be moved either virtually or physically), two multiple access solutions have been proposed: \textit{i)} fast-\gls{FAMA}, where the system switches on a symbol-by-symbol basis \cite{Khammassi2023, Wong2021}; and \textit{ii)} slow-\gls{FAMA}, where switching occurs only when the channel changes (i.e., every time-coherence interval) \cite{Wong2022}. While the former is deemed impractical due to implementation limitations, the latter arises as a more feasible and realistic solution. However, despite the promising potential of FAMA systems, their true multiple access capabilities are still not fully understood, particularly when considering 2D surfaces. The promising preliminary results on 2D FAS \cite{Kiat2023}, along with the advent of reconfigurable intelligent surfaces (RIS), suggest that 2D FAMA could indeed be a key enabler for massive multiple access in increasingly dense 6G networks. Research to that end is still at an early stage and further studies are needed, both on physical realizations and theoretical analysis (modeling and performance).

A key limitation of \gls{FAS} is that the channel in the space continuum is \textit{inherently correlated}. Having (infinitely) many switchable antenna ports (pixels, or switching positions) packed, adjacent to one another, unavoidably increases the correlation between these positions. A well-known fundamental fact is that, to sample ``independent" channels, antennas should be placed at a minimum distance of $\lambda/2$ \cite{Pizzo2020, Wong2021_FAS}.
%; a theoretical result that arises from spatial correlation in isotropic propagation environments \cite{Pizzo2020, Wong2021_FAS}.
This could be seen as the ``spatial sampling rate" to exploit the channel's spatial diversity, raising the fundamental question: \textit{what is the benefit of the ``oversampling" in \gls{FAS}?} Indeed, the correlation between antenna ports (sampling points) increases as more ports are packed within a given aperture, i.e., as we oversample the aperture.
%Besides, the larger the number of ports for a given aperture---spatial points where the received signal can be sampled---the higher the correlation between them.
This renders diminishing returns and, ultimately, leads to a saturation effect as shown in \cite{Khammassi2023, Kiat2023_New}. Hence, \textit{how much oversampling do we need to realize the full potential of \gls{FAS}?} These two fundamental questions remain still open. 

Naturally, to properly address the previous questions and to theoretically assess the performance of \gls{FAS}, appropriate spatial correlation models need to be considered. Thus far, \rev{although some works dive into general (geometric-based) channel formulations \cite{Zhu2023Model, Ma2024},} most of the works on \gls{FAS} widely adopt the classical Jakes's correlation model (e.g., \cite{Wong2022, Wong2021, Khammassi2023, Wong2022_EL, Tang2023, Wong2021_FAS, Zhang2023, Skouroumounis2023, Kiat2023, Kiat2023_New, Psomas2023}), well founded from a physics viewpoint, assuming 2D isotropic propagation with scatterers uniformly distributed in a ring around the receiver \cite{Jakes1994}. However, the analysis of \gls{FAS} under Jakes's model is prohibitively complex. Substantial efforts have been made to come up with simplified, yet accurate correlation models, aimed to serve as good approximations while allowing for a tractable analysis \cite{Wong2022_EL, Wong2022, Khammassi2023, Kiat2023_New}. An approximation to Jakes's correlation---based on a few dominant eigenvalues of the correlation matrix---is proposed in \cite{Khammassi2023, Kiat2023_New}, resulting in intricate expressions with large numbers of nested integrals---as many as the number of \gls{FAS} antenna ports. A simpler model (analysis is possible) was introduced in \cite{Wong2022_EL}, where Jakes's correlation is replaced by a ``constant" correlation parameter such that pairs of ports are equally correlated.
%rendering a correlation matrix with equal entries. 
Despite having been assumed since early \gls{FAS} contributions  \cite{Vega2023, Wong2022, Tlebaldiyeva2023, Yang2023, Lai2023,Vega2023_Simple}, its validity to realistically (accurately) model spatial correlation is questioned \cite{Khammassi2023} (as we shall also see later). Indeed, the oversimplified ``constant" correlation may often yield misleading conclusions on the performance of FAS.

Motivated by the outstanding need of appropriate (simple and accurate) models, we here propose a new method to characterize the spatial correlation in \gls{FAS}. Our goal is to accurately model correlation, as predicted by classical realistic models such as Jakes's, while retaining the analytical tractability and simplicity of the ``constant" correlation model in \cite{Wong2022_EL}.
%we here propose a new method to characterize the spatial correlation in \gls{FAS}.
Inspired by the coherence interval idea behind block-fading models\footnote{\rev{The reader is gently referred to standard textbooks on block-fading channels \cite[ch. 4]{Goldsmith2005}\cite[ch. 2]{Tse2005}.}}, we consider that the spatial correlation remains approximately constant within a block (set of ports), while different blocks are assumed independent. This resembles the concept of time coherence interval in Jakes's channel autocorrelation model.
Based on this \emph{spatial coherence interval}, the key idea is approximating a given spatial correlation matrix (e.g., Jakes's) with a block-diagonal matrix by means of spectral analysis.
Statistical theory for large Toeplitz matrices (by extension, correlation matrices) dictates that, as the number of ports increases for a given aperture, the matrix tends to be dominated by a small fraction of eigenvalues (as noted in \cite{Khammassi2023, Kiat2023_New}). Hence, each of the blocks can ``capture" one of the eigenvalues, yielding a block-diagonal matrix with similar spectrum to that of the target correlation matrix. This approximation framework enables a tractable analysis of \gls{FAS} and paves the way towards further studies on their fundamental limits. In summary, the contributions of this paper are:
\begin{itemize}
    \item We propose a block-diagonal spatial correlation model for \gls{FAS} that provides a trade-off solution between accuracy and mathematical tractability. The model can capture the spectral characteristics of any arbitrary correlation function; it is therefore generally applicable to approximate any realistic spatial correlation structure beyond the classic ones (e.g., Jakes's and Clarke's).  
    \item Building upon statistical theory of large Toeplitz matrices, we provide an algorithm to efficiently approximate the \gls{FAS} correlation structure based on the block-diagonal framework. The proposed method is particularly simple to evaluate, in contrast to conventional correlation models (especially for 2D fluid antennas).
    \item The proposed solution is applied to slow-\gls{FAMA} systems, yielding tractable expressions for the \gls{OP}, as well as simple approximations and upper bounds.
    \item A performance evaluation of slow-\gls{FAMA} is carried out, highlighting the benefit of the oversampling rendered by \gls{FAS}, its multiplexing capacity, and the ability of the block-correlation model to capture the different effects (e.g., gain saturation effect) inherent to FAS.
\end{itemize}

The remainder of this paper is organized as follows: Section \ref{sec:FAS_channel} provides an overview of spatial correlation models and presents the channel model for \gls{FAS}. Section \ref{sec:BlockApprox} details the proposed framework to approximate correlation in \gls{FAS}. Section \ref{sec:Slow-FAMA} carries out the performance analysis of slow-\gls{FAMA} systems based on the block-diagonal approximation. Section \ref{sec:Numerical} presents the performance evaluation of slow-\gls{FAMA} and, finally, some conclusions are drawn in Section \ref{sec:Conclusions}.

\textit{Notation:} Vectors and matrices are represented by bold lowercase and uppercase symbols, respectively, $(\cdot)^T$ denotes the matrix transpose, and $\|\cdot\|_2$ is the $\ell_2$ norm of a vector. Also, $(\mat{A})_{j,k}$ is the $j,k$-th element of $\mat{A}$. Finally, $j = \sqrt{-1}$ is the imaginary number, and $\mathbb{E}[\cdot]$ is the mathematical expectation. Any other specific notation will be defined when necessary.

\rev{\textit{Supplementary files: } The \textsc{Matlab} code used for simulations and figures in this paper can be found at \url{https://github.com/preugr/Fluid-antenna-block-diagonal-model.git}.}

%------------------------------------------------------------------------------------------------------------------
% CHANNEL AND SPATIAL CORRELATION
%------------------------------------------------------------------------------------------------------------------
\section{Spatial Correlation and Channel Models}

\label{sec:FAS_channel}

In principle, channel modeling for \gls{FAS} should not be different from classical multi-antenna channel modeling, with one key exception: the capital importance of \emph{spatial correlation}, as a consequence of the arbitrarily close spacing between antenna ports.
We here revisit a general channel modeling approach and how the different assumptions on the propagation environment lead to different correlation models. 

\subsection{Spatial correlation for Rayleigh channels}

\label{sec:SpatialCorrelation}

\begin{figure}[t]
    \centering
    \includegraphics[width =0.5\columnwidth]{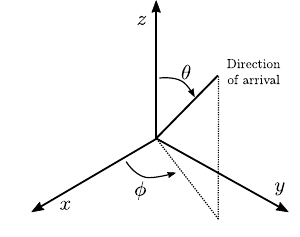}
    \caption{Coordinates system.}
    \label{fig:Coordinates}
\end{figure}

Consider, as in \cite{Aulin1979}, an antenna element at an arbitrary position $\vec{r}_n = \begin{pmatrix} x_n & y_n & z_n\end{pmatrix}^T$ in the 3D space. Assuming far-field propagation and some source (transmitter) exciting the scenario with a narrowband transmission, the complex base-band signal (channel) $h_n$ received at $\vec{r}_n$ is given by the superposition of $P$ plane waves (paths) as
\begin{equation}
    h_n = \sum_{p=1}^P \alpha_p \,G(\phi_p,\theta_p)\,\exp{j\vec{k}(\phi_p,\theta_p)^T\vec{r}_n}, \label{eq:h_general}
\end{equation}
where $\alpha_p\in\mathbb{C}$ is the gain of the $p$-th path, $\phi_p\in[0, 2\pi)$ and $\theta_p\in[0, \pi)$ are the azimuth and polar arrival angles (see Fig. \ref{fig:Coordinates}), $G(\phi_p,\theta_p)$ is the antenna pattern, and 
\begin{equation}
    \vec{k}(\phi_p,\theta_p) = k\begin{pmatrix} \sin{\theta_p} \cos \phi_p & \sin\theta_p \sin \phi_p & \cos\theta_p\end{pmatrix}^T
\end{equation}
with $k=2\pi/\lambda$ the wavenumber. To further elaborate, we need to impose conditions on some of the variables in \eqref{eq:h_general}. A common assumption is that $\alpha_p$ $\forall$ $p$ are independent random variables\footnote{The randomness is justified by assuming that each plane wave in \eqref{eq:h_general} is the result of multiple unresolvable reflections within the %scatterer rendering the $p$-th path.} 
scatterer.}
with $\mathbb{E}[\alpha_p] = 0$ and $\mathbb{E}[|\alpha_p |^2] = \sigma_\alpha^2/P$, with $\sigma_\alpha^2>0$. That is, the total received power is equally distributed between all the incoming plane-waves. Besides, unless going very high in frequency, the number of paths $P$ is typically large, and thus the central limit theorem applies. Thus, $h_n$ is Gaussian with %distributed with moments
\begin{align}
     \mathbb{E}[h_n] &\overset{(a)}{=} \sum_{p=1}^P \mathbb{E}[\alpha_p] \,\mathbb{E}[G(\phi_p,\theta_p)\,e^{j\vec{k}(\phi_p,\theta_p)^T\vec{r}_n}] = 0, \\
      \mathbb{E}[|h_n|^2] &\overset{(a)}{=} \frac{\sigma_\alpha^2}{P}\sum_{p=1}^P\mathbb{E}[|G(\phi_p,\theta_p)|^2],
\end{align}
where the independence between $\alpha_p$ and $\theta_p$, $\phi_p$ is assumed in $(a)$. This twofold assumption gives rise to the so-called Rayleigh fading. We are interested in the relation between the channel experienced by two nearby antenna elements located at $\vec{r}_n$ and $\vec{r}_m$, which, by the plane wave assumption, experience the same path gains and angles of arrival. Thus,
\begin{align}
    \sigma_{n,m}^2 &= \mathbb{E}[h_n h_m^*] \notag \\
    &= \frac{\sigma_\alpha^2}{P}\sum_{p=1}^P\mathbb{E}[|G(\phi_p,\theta_p)|^2\,\exp{j\vec{k}(\phi_p,\theta_p)^T(\vec{r}_n-\vec{r}_m)}] \notag \\
    &\overset{(b)}{=} \sigma_\alpha^2 \,\mathbb{E}[|G(\phi,\theta)|^2\,\exp{j\vec{k}(\phi,\theta)^T(\vec{r}_n-\vec{r}_m)}], \label{eq:Sigma_General}
\end{align}
where $(b)$ follows after assuming that, for every path, the angles of arrival are independent and identically distributed (i.i.d.) with density $f_{\theta,\phi}(\theta,\phi)$.

The spatial correlation for the Rayleigh channel depends therefore on \textit{i)} the radiation pattern of the antennas, and \textit{ii)} the distribution of the angles of arrival, imposed by the propagation environment. For simplicity, isotropic antennas are widely assumed, i.e., $|G(\phi,\theta)|^2=1$. Isotropic propagation is also commonly assumed, giving rise to the 3D Clarke's model \cite{Aulin1979}. In this case, $\theta$ and $\phi$ are independent with densities
\begin{align}
    f_\theta(\theta) &= \frac{\sin(\theta)}{2}, & f_\phi(\phi) &= \frac{1}{2\pi},  \label{eq:3DClarkeDistro}
\end{align}
which, introduced into \eqref{eq:Sigma_General}, lead to \cite{Aulin1979, Bjornson2021, Pizzo2020} 
\begin{equation}
    \sigma_{n,m}^2 = \sigma_\alpha^2\frac{\sin\left(k \|\vec{r}_n-\vec{r}_m\|_2\right)}{k \|\vec{r}_n-\vec{r}_m\|_2} = \sigma_\alpha^2 \text{sinc}\left(k \|\vec{r}_n-\vec{r}_m\|_2\right), \label{eq:3DClarke}
\end{equation}
where $\text{sinc}(z) = \frac{\sin(z)}{z}$. Clearly, we see that $\sigma_{n,m}^2 = 0$ for $\|\vec{r}_n-\vec{r}_m\|_2 = \lambda/2$, which gives the well-known rule of thumb that half-wavelength spacing yields independent channels. 

Consider now a 2D scenario, i.e., $\vec{r}_n = (x_n \,\,\, y_n)$, which corresponds to fixing $\theta = \pi/2$. Assume the same distribution for the azimuth angle as in 3D Clarke's model, given by $f_\phi(\phi) = \frac{1}{2\pi}$, which ultimately implies that all the plane waves arrive uniformly distributed in a ring around the receiver. Introducing these considerations into \eqref{eq:3DClarkeDistro} yields now
\begin{equation}
    \sigma_{n,m}^2 = \sigma^2_\alpha J_0\left(k\|\vec{r}_n-\vec{r}_m\|_2\right), \label{eq:Jakes}
\end{equation}
where $J_\nu(\cdot)$ is the $\nu$-th order Bessel function of the first kind. Note that \eqref{eq:Jakes} corresponds to Jakes's correlation function (2D Clarke's model) \cite{Jakes1994, Goldsmith2005, Tse2005}. 

\begin{remark}
The two examples above characterize the spatial channel correlation under isotropic propagation conditions. For an arbitrary scenario, the correlation function can be directly calculated by introducing the corresponding distributions for the angles of arrival and antenna radiation pattern in \eqref{eq:Sigma_General}.
\end{remark}

\subsection{Channel model for FAS}
\label{subsec:FAS_channel_model}

From a channel modeling perspective, a \gls{FAS} can be seen as a collection of colocated radiating elements, representing the ports where the fluid antenna can switch into. Denoting by $N$ the number of ports, the channel vector is expressed as $\vec{h} = \begin{pmatrix}h_1(\vec{r}_1) & \dots & h_N(\vec{r}_N) \end{pmatrix}$, where the dependence with the (3D) position is explicitly stated. According to Rayleigh fading, as seen before, $\vec{h}$ is jointly Gaussian, i.e., $\vec{h}\sim\mathcal{CN}_N(\vec{0}, \sigma_\alpha^2 \bm{\Sigma})$, with $\bm{\Sigma}\in\mathbb{C}^{N\times N}$ the spatial correlation matrix. As discussed previously, the correlation matrix depends on the propagation environment, being in general described by $(\bm{\Sigma})_{n,m} = \sigma_{n,m}^2/\sigma_\alpha^2$. Naturally, the structure of $\bm{\Sigma}$ depends on the fluid antenna topology and, specifically, on the ports positions $\vec{r}_1,\dots, \vec{r}_N$. Thus, in a 1D fluid antenna, $\bm{\Sigma}$ has a Toeplitz structure. In the 2D case, $\vec{h}$ represents the vectorized channel matrix, and the structure of $\bm{\Sigma}$ is imposed by the port numbering, e.g., if the ports are numbered row-by-row, $\bm{\Sigma}$ is block Toeplitz. Note however that the physical structure is the same regardless of the port arrangement. 

\begin{figure}[t]
    \centering
    % Trim option due to figure saved directly from Matlab
    \includegraphics[trim = {0 7.5cm 0 7.5cm}, clip, width = \columnwidth]{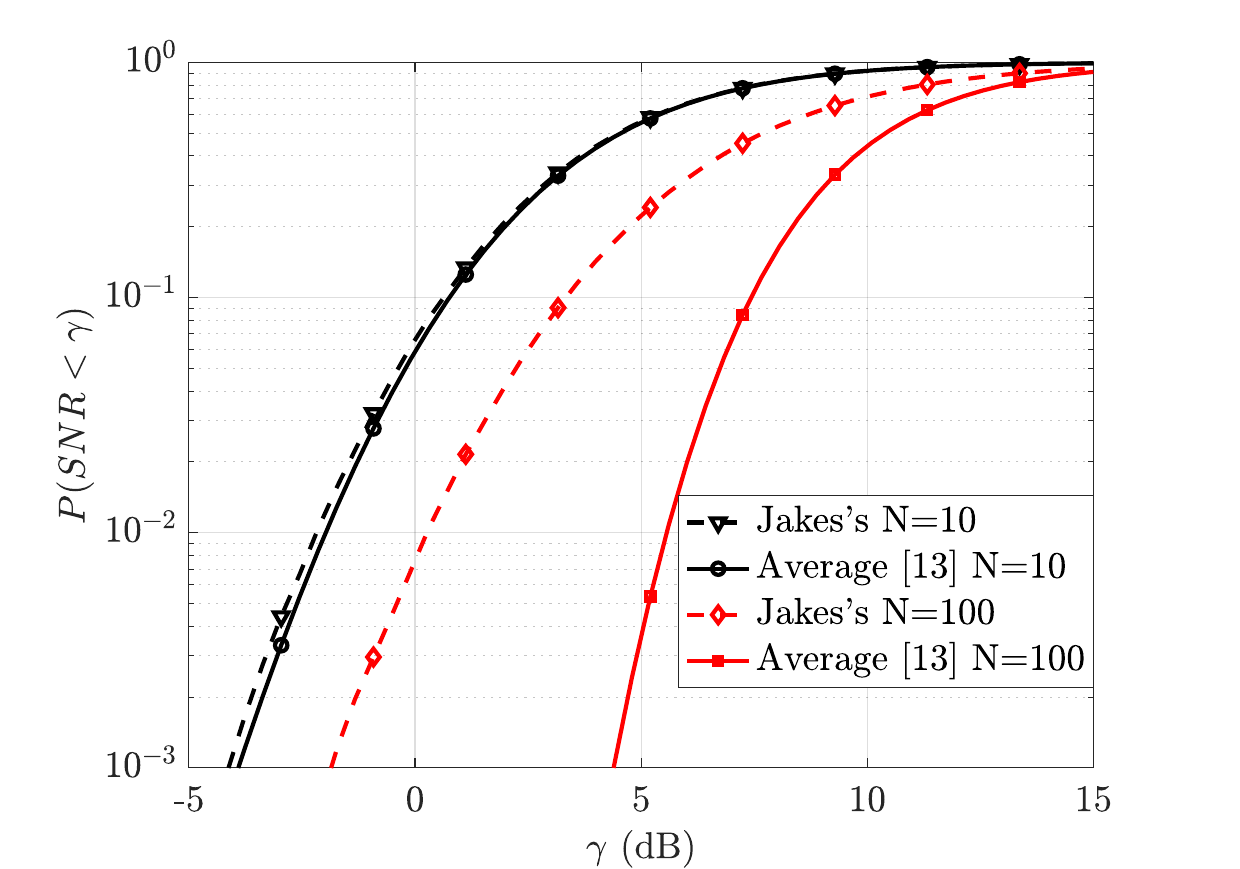}
    \caption{Outage probability for 3 users when using a linear fluid antenna with length $4\lambda$ under Jakes's correlation model and the ``average'' model in \cite{Wong2022_EL}.}
    \label{fig:Example_Avg_Jakes}
\end{figure}

Jakes's correlation is widely adopted in the literature for 1D fluid antennas; thus, $\sigma_{n,m}^2$ is given by \eqref{eq:Jakes} and
\begin{equation}
    (\bm{\Sigma})_{n,m} = J_0\left(k\|\vec{r}_n-\vec{r}_m\|_2\right), \label{eq:Sigma_Jakes}
\end{equation}
representing isotropic propagation scenarios as validated in \cite[Chapter 1]{Jakes1994}. However, note that Jakes's model is not valid for planar \gls{FAS}, since it assumes a 2D propagation environment. Unfortunately, analytical characterization of \gls{FAS} under Jakes's or Clarke's (for planar \gls{FAS}) models is prohibitively complex. A twofold approximation to Jakes's correlation is proposed in \cite{Khammassi2023}; the first approximation selects only the dominant eigenvalues of $\bm{\Sigma}$ (\rev{a similar rank-reduction approximation is used in \cite{Kiat2023_New}}), while the second one aims to approximate the resulting matrix by another one with similar distribution but more amenable to analysis. Still, the resulting analysis is intricate. A much simpler approach is presented in \cite{Wong2022_EL}, modeling the channel at each port as a correlated version of a reference (and common) variable as in \cite{Beaulieu2011}, i.e., $h_n = \sqrt{1-\mu_n^2}x_n + \mu_n x_0$, where $x_n, x_0\sim\mathcal{CN}(0,1)$. To minimize the number of parameters, \cite{Wong2022_EL} proposes to use a unique $\mu_n = \mu$ $\forall$ $n$, chosen as the average correlation along the fluid antenna, yielding the constant-correlation matrix
\begin{equation}
    \bm{\Sigma}_\text{avg} = \begin{pmatrix} 1 & \mu^2 & \mu^2 &\dots & \mu^2 \\ 
    \mu^2 & 1 & \mu^2 & \dots & \mu^2 \\
    \vdots & & \ddots & & \vdots \\
    \mu^2 & \dots & \mu^2 & \mu^2 & 1\end{pmatrix}, \label{eq:Sigma_Kit}
\end{equation}
which considerably improves the tractability of the model yielding, e.g., relatively simple expressions for the \gls{OP} for \gls{FAMA} systems. This simplicity comes at the price of a possibly poor accuracy,
%when approximating \eqref{eq:Sigma_Jakes}, 
as illustrated in Fig. \ref{fig:Example_Avg_Jakes}, where we see the \gls{OP} of a \gls{FAMA} system under the average (constant) correlation model of  \eqref{eq:Sigma_Kit} \cite{Wong2022_EL}, and under Jakes's model. Observe that, for a reduced number of ports, the model in \eqref{eq:Sigma_Kit} is accurate (mainly, because a low number of ports $N$ translates into large spacing and hence low correlation), but the gap (inaccuracy) considerably increases as the fluid antenna is densified---precisely the objective of \gls{FAS}. Thus, as $N$ becomes large for a given aperture, the performance of \gls{FAS} seems to be dominated by the high correlation between close-by ports, and conclusions drawn from the said constant-correlation model should at the very least be questioned. 

In summary, existing modeling approaches are either too complex from an analytic viewpoint (the classical realistic models like Jakes's or Clarke's) or poorly accurate because of their over-simplicity (tractable approximations like the models in \cite{Wong2022_EL,Wong2022}). There is therefore an outstanding need for alternative, simple and yet accurate spatial correlation models. Aiming to address this need, we next propose a new spatial correlation model, based on a block-diagonal approximation.

%------------------------------------------------------------------------------------------------------------------
% BLOCK DIAGONAL APPROXIMATION
%------------------------------------------------------------------------------------------------------------------
\section{Spatial Block-Correlation Model}

\label{sec:BlockApprox}

With our proposed model, we aim to approximate a given (realistic) spatial correlation structure (such as Jakes's) by block-diagonal matrices. The rationale behind this modeling choice is based on the similarity (parallelism) between the channel's dynamic behaviour in time and in space. In the time domain, the well-known block-fading approximation assumes independent temporal blocks with constant channel values within each block \rev{\cite[ch. 4]{Goldsmith2005}}, where the block length is determined by the channel coherence interval. Inspired by this, we here explore the idea of a block-correlation approximation to the channel's spatial autocorrelation. That is, we aim to characterize the spatial correlation in the fluid antenna by a block-diagonal matrix, where the correlation within each block is constant (as in temporal block fading) but the different blocks are independent. To that end, we first look into the spectral characteristics of the target spatial correlation.

%The correlation models presented previously---Jakes's and Clarke's---were originally developed to characterize the temporal autocorrelation of the channel; it is through the ergodicity assumption that they are extended to spatial correlation processes. 

\subsection{Spectral analysis of spatial correlation matrices in FAS}
\label{sec:SpectralAnalysis}

The entries of the spatial correlation matrix $\bm{\Sigma}$ are obtained by ``sampling" a continuous correlation function, e.g., \eqref{eq:3DClarke} or \eqref{eq:Jakes}, at points spaced by the distance between the $N$ ports. The key difference with respect to conventional antenna systems is that $N$ is very large and, thus, the correlation function is densely sampled within the fluid antenna aperture. Leveraging sampling theory, this means we are \textit{oversampling} the correlation function, and hence many samples are redundant and provide no extra information. Intuitively, we may think that this translates into the correlation matrix $\bm{\Sigma}$ (and its spectral characteristics) and, as $N$ becomes large, the extra entries of $\bm{\Sigma}$ provide no additional knowledge on the correlation, leading to a rank-deficient matrix. In other words, as $N$ increases, we could expect $\bm{\Sigma}$ to be dominated by a few eigenvalues. %being that more evident . 

To get further insight, consider the case of a linear (1D) fluid antenna and a 3D isotropic environment such that the spatial correlation is given by \eqref{eq:3DClarke}. Denoting by $W$ the length of the fluid antenna normalized by the wavelength, and assuming that the ports are equally spaced, $\bm{\Sigma}$ is given by the Toeplitz matrix 
\begin{align}
    \bm{\Sigma} &= \left(\begin{smallmatrix}
        g(0) & g(1) & g(2) & \dots & g(N-1) \\
        g(-1) & g(0) & g(1) & \dots & g(N-2) \\
        \vdots &    & \ddots &     & \vdots \\
        g(-N+1) & g(-N+2) & \dots & g(-1) & g(0)
    \end{smallmatrix}\right), \label{eq:Sigma_Clarke3D} 
\end{align}
where the generating function is 
\begin{equation}
    g(n) = \text{sinc}\left(\frac{2\pi n W}{N-1}\right). 
\end{equation}
Building upon statistical results on large Toeplitz matrices, the asymptotic spectral properties of $\bm{\Sigma}$ are studied next.

%\begin{lemma}
%    \label{lemma:3DClarke} Denote by $\rho_n$ for $n = 1,\dots,N$ the eigenvalues of $\bm{\Sigma}$ in \eqref{eq:Sigma_Clarke3D}, and consider the truncated Fourier series 
%    \begin{equation}
%        f_N(x) = \sum_{n=-N/2+1}^{N/2-1} g(n) e^{jnx}, \quad\quad x\in[-\pi,\pi). \label{eq:FourierSeries}
%    \end{equation}
%    Then, the sequences $\{\rho_n\}_{n=1}^N$ and $\left\{f_N\left(\frac{2\pi (n-1)}{N}\right)\right\}_{n=1}^N$ are equally distributed in the limit $N\rightarrow\infty$, should $f_N(x)$ be square-summable \cite[Lemma 4.2 and Theorem 5.1]{Tyrtyshnikov1996}.
%\end{lemma}

%\begin{corollary}
%    Consider an arbitrary small threshold $\rho_\text{th}$ not larger than $\frac{N-1}{2W}$. Then, as $N\rightarrow \infty$ for fixed $W$, the number of relevant eigenvalues of $\bm{\Sigma}$ in \eqref{eq:Sigma_Clarke3D} surpassing $\rho_\text{th}$ is approximated by $2W$. \label{coro:Clarke3D}
%\end{corollary}

\rev{\begin{lemma}
    \label{lemma:Clarke3D}
    Denote by $\rho_n$ for $n = 1, \dots, N$ the eigenvalues of $\bm{\Sigma}$ in \eqref{eq:Sigma_Clarke3D}, and consider an arbitrary small threshold $\rho_\text{th}$ not larger than $\frac{N-1}{2W}$. Then, as $N\rightarrow \infty$ for fixed $W$, the number of relevant eigenvalues of $\bm{\Sigma}$ surpassing $\rho_\text{th}$, i.e., $|\{\rho_n \,|\,\rho_n > \rho_\text{th}\}|$ is approximated by $2W$.
\end{lemma}
\begin{IEEEproof}
    From \cite[Lemma 4.2 and Theorem 5.1]{Tyrtyshnikov1996}, we have that, for any continuous function $F(x)$, then
    \begin{equation}
        \lim_{N\rightarrow \infty} \frac{1}{N}\sum_{n=1}^N F(\rho_n) = \frac{1}{2\pi}\int_{-\pi}^\pi F\left(\lim_{N\rightarrow\infty}f_N(x)\right) \dif x, \label{eq:EquallyDistribution}
    \end{equation}
    where $f_N(x)$ is the square-summable truncated Fourier series
    \begin{equation}
        f_N(x) = \sum_{n=-N/2+1}^{N/2-1} g(n) e^{jnx}, \quad\quad x\in[-\pi,\pi). \label{eq:FourierSeries}
    \end{equation}

    Choosing $F(x)$ as an approximation to the step function\footnote{Note that the step function itself is not a valid function since it violates the continuity assumption.}, the portion of eigenvalues larger than $\rho_\text{th}$ can hence be approximated by the right-hand side of \eqref{eq:EquallyDistribution}. In the case of \eqref{eq:Sigma_Clarke3D}, it can be checked that the Fourier series in \eqref{eq:FourierSeries} corresponds to a rectangular wave, i.e., for large $N$ we have
    \begin{align}
        f_N(x) &= \sum_{n=-N/2+1}^{N/2-1} \text{sinc}\left(\frac{2\pi n W}{N-1}\right)e^{jnx}\notag \\ &\approx \begin{cases} \frac{N-1}{2W} & \text{if } |x| \leq \frac{2\pi W}{N-1} \\ 0 & \text{otherwise} \end{cases}.
    \end{align}

    Introducing the above result in \eqref{eq:EquallyDistribution}, and using the logistic function as the step function approximation, i.e., $F(x) = \left(1+e^{-\beta (x - \rho_\text{th})}\right)^{-1}$ with $\beta>0$, we readily obtain (with a slight abuse of notation)
    \begin{align}
        \frac{1}{2\pi}\int_{-\pi}^\pi &F\left(\lim_{N\rightarrow\infty}f_N(x)\right) \dif x \approx \left(1-\frac{2W}{N-1}\right)\frac{1}{1+e^{\beta\rho_\text{th}}} \notag \\
        &+\frac{2W}{N-1}\frac{1}{1+e^{-\beta((N-1)/(2W) - \rho_\text{th})}}.
    \end{align}
    The proportion of eigenvalues surpassing $\rho_\text{th}$ is finally obtained by setting $\beta$ large (so that the logistic function approximates the unitary step), leading to $\frac{2W}{N-1}$, and therefore the number of relevant eigenvalues is $\frac{2W}{N-1}N\approx 2W$.
\end{IEEEproof}}

\rev{Lemma \ref{lemma:Clarke3D}} nicely connects with sampling theory, stating that the number of dominant eigenvalues (samples)\rev{---eigenvalues larger than a threshold $\rho_\text{th}$ less than $\frac{N-1}{2W}$---}is approximately the number of half-wavelengths contained in the fluid antenna aperture. \rev{Note, though, that for large $N$ the threshold $\rho_\text{th}$ is large, and hence $\bm{\Sigma}$ is dominated by approximately $2W$ large eigenvalues, being the rest negligible}. The same analysis can be carried out for Jakes's correlation \eqref{eq:Sigma_Jakes}, as shown next.

\begin{corollary}
    Assume a 1D fluid antenna of normalized size $W$, where the spatial correlation $\bm{\Sigma}$ is given by the Toeplitz matrix generated by $g(n) = J_0(\frac{2\pi n W}{N-1})$ (Jakes's correlation). Then, as $N\rightarrow \infty$, the number of eigenvalues exceeding a relatively small threshold ($\rho_\text{th} \hspace{-0.05cm}< \hspace{-0.05cm} \frac{N-1}{2\pi W}$) is approximated by $2W$. \label{coro:Jakes}
\end{corollary}
\begin{IEEEproof}
    With $g(n) = J_0\left(\frac{2\pi n W}{N-1}\right)$ in \eqref{eq:FourierSeries} and using the transform pair in \cite[p. 122, Eq. (2)]{Bateman1954}, we have, for large $N$,
\begin{align}
    f_N(x)  \approx \begin{cases} \frac{N-1}{W\pi}\frac{1}{\sqrt{1-\left(\frac{N-1}{2\pi W} x\right)^2}} & \text{if } |x| \leq \frac{2\pi W}{N-1} \\ 0 & \text{otherwise} \end{cases}.
\end{align}
    The proof then follows the same steps of Lemma \ref{lemma:Clarke3D}, using again the logistic approximation to the step function.
\end{IEEEproof}

\begin{figure}[t]
    \centering
    % Trim option due to figure saved directly from Matlab
    \includegraphics[trim = {0 7.5cm 0 7.5cm}, clip,width = \columnwidth]{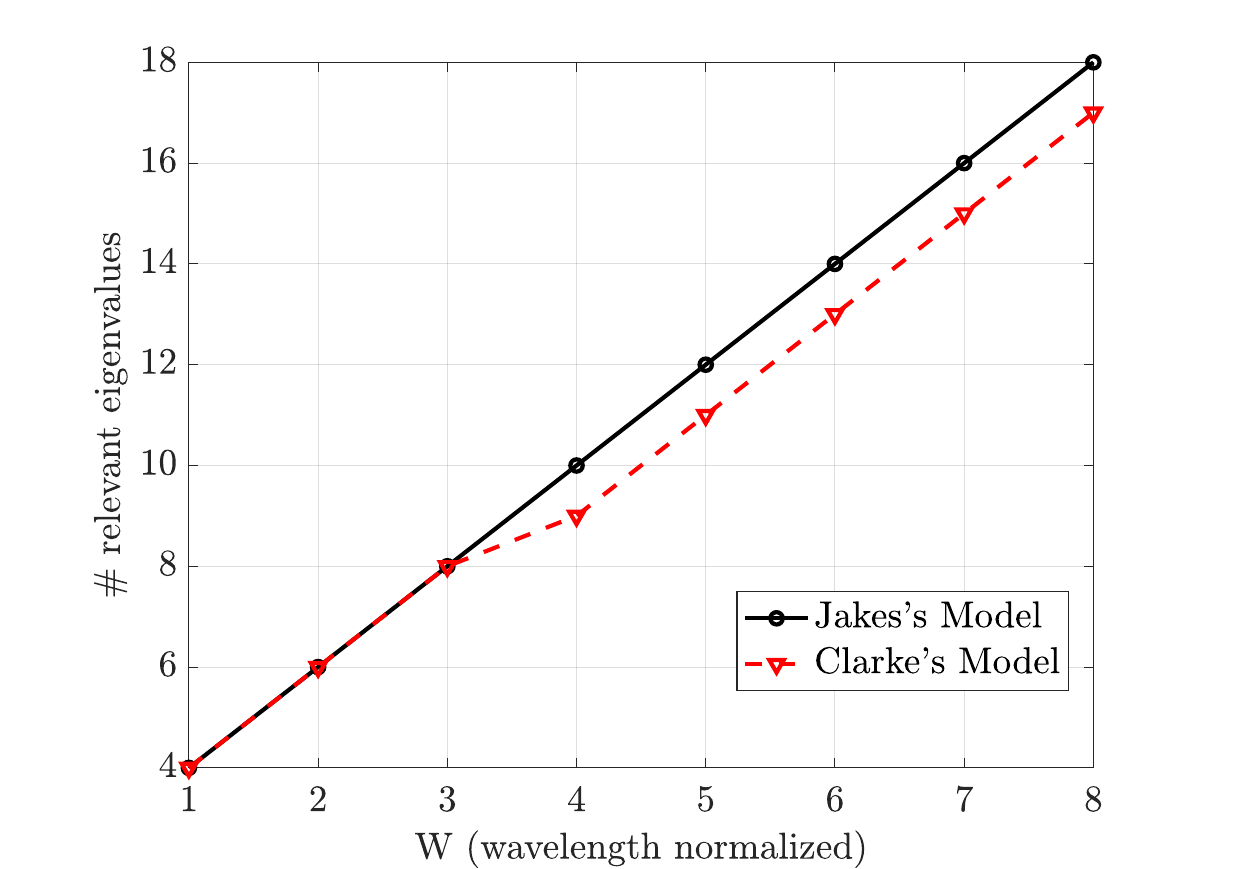}
    \caption{Number of relevant eigenvalues vs. fluid antenna length (normalized by the wavelength); 20 ports per wavelength.}
    \label{fig:Relevant_eigenvalues}
\end{figure}

Note that the result in Corollary \ref{coro:Jakes} was obtained in \cite{Khammassi2023}. As an illustrative example, the number of relevant eigenvalues larger than $N/100$, i.e., containing $99$\% of the ``power", is depicted in Fig. \ref{fig:Relevant_eigenvalues}, where the linear relation with $W$ is clearly observed. Notice, however, that the number of eigenvalues is not exactly $2W$, this being an asymptotic result, and because of the arbitrary threshold. 

\begin{figure}[t]
    \centering
    % Trim option due to figure saved directly from Matlab
    \includegraphics[trim = {0 7.5cm 0 7.5cm}, clip, width = \columnwidth]{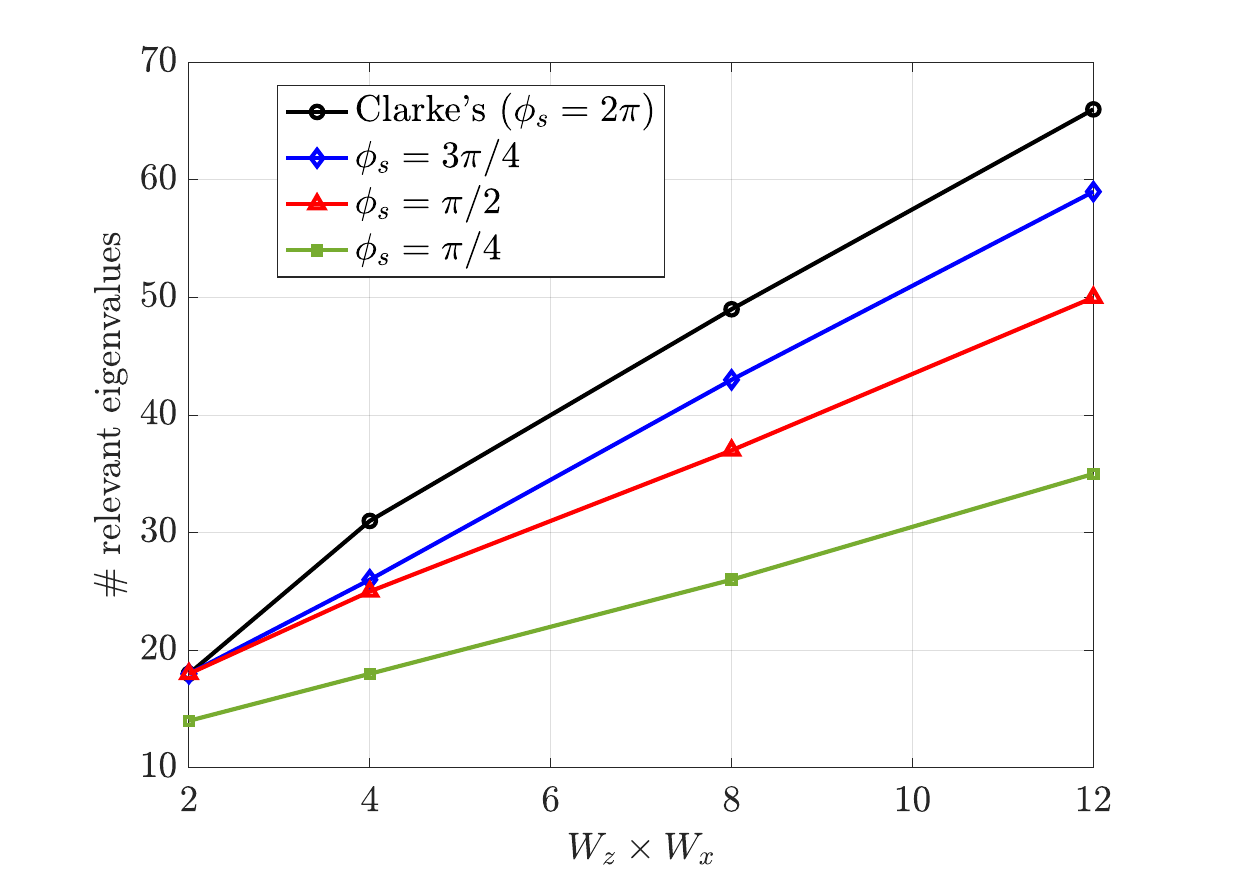}
    \caption{\rev{Number of relevant eigenvalues vs. fluid antenna area for different azimuth spans; the sizes $W_z$ and $W_x$ are normalized by the wavelength; 100 ports per unit area.}}
    \label{fig:Relevant_eigenvalues_2D}
\end{figure}

In the case of a 2D fluid antenna, since the resulting correlation matrix is no longer Toeplitz, the above statistical results do not directly apply. However, we may expect a similar trend, with the spatial oversampling leading to a rank-defficient correlation matrix. \rev{This is observed in Fig. \ref{fig:Relevant_eigenvalues_2D}, which plots the relevant eigenvalues (larger than 1) of a 2D fluid antenna of wavelength-normalized sizes of $W_z\times W_x$ under different correlation models. Again, we see a clear linear relation between the number of dominant eigenvalues and the antenna size (in this case, its area), but this relation is not the same for all the correlation models. More specifically, the isotropic propagation assumed in Clarke's model yields the largest degrees of freedom (dominant eigenvalues in $\bm{\Sigma}$), since the azimuth domain is the whole space ($2\pi$) as seen in \eqref{eq:3DClarkeDistro}. Naturally, as we restrict the angular domain, the incoming plane waves arrive from more similar directions, and hence the spatial correlation increases, leading to less degrees of freedom in the channel (equivalently, less dominant eigenvalues). The different correlation models in Fig. \ref{fig:Relevant_eigenvalues_2D} are directly computed from \eqref{eq:Sigma_General} but restricting the azimuth to be uniformly distributed in $[{\pi}/{2}-{\phi_s}/{2}, {\pi}/{2}+{\phi_s}/{2}]$ (the polar angle distribution remains unchanged). Clearly, $\phi_s = 2\pi$ corresponds to Clarke's model. Albeit interesting, a detailed analysis of the impact of the propagation environment in the correlation matrix---and consequently the performance of \gls{FAS}---is out of the scope of this paper due to space constraints, and thus we stick to Jakes's and Clarke's models in the following.}

Thus, building upon statistical results on large Toeplitz matrices, we have seen that both Clarke's 3D and Jakes's correlation models lead to correlation matrices which, when oversampled, are dominated by a few eigenvalues (specifically, the number of half-wavelengths contained in the fluid antenna aperture). This result will be used in the next subsection to justify a block-diagonal approximation to $\bm{\Sigma}$.

\subsection{Block-diagonal matrix approximation}
\label{sec:Blockapprox}

Motivated by the tractability of the constant correlation matrix in \eqref{eq:Sigma_Kit} \cite{Wong2022}, and by the fact that $\bm{\Sigma}$ is dominated by a very few eigenvalues, we propose a block-diagonal approximation to the true (target) correlation matrix with similar spectrum. That is, we seek for a matrix of the form 
\begin{equation}
    \widehat{\bm{\Sigma}} \in\mathbb{R}^{N\times N}= \begin{pmatrix}
    \mat{A}_1 & \mat{0} & \cdots & \mat{0} \\
    \mat{0} & \mat{A}_2 & \cdots & \mat{0} \\
    \vdots & & \ddots & \vdots \\
    \mat{0} & \mat{0} & \mat{0} & \mat{A}_B
    \end{pmatrix},    \label{eq:SigmaHat}
\end{equation}
where each submatrix $\mat{A}_b$ is a constant correlation matrix of size $L_b$ and correlation $\mu_b^2$, i.e.,
\begin{equation}
    \mat{A}_b \in\mathbb{R}^{L_b\times L_b} = \begin{pmatrix}
    1 & \mu_b^2 & \cdots & \mu_b^2 \\
    \mu_b^2 & 1 & \cdots & \mu_b^2 \\
    \vdots & & \ddots & \vdots \\
    \mu_b^2 & \cdots & \mu_b^2 & 1
    \end{pmatrix},  \quad b = 1,\dots,B.
\end{equation}
We are then interested in solving the general problem 
\begin{equation} 
 \mathop {\text{arg min}}\limits_{B,\;L_1,\dots,L_B,\;\mu_1,\dots,\mu_B} \;\;\mathop \text{dist}\left(\widehat{\bm{\Sigma}}, \bm{\Sigma} \right)
\end{equation}
where $\text{dist}(\cdot)$ is a distance metric between the approximated and true correlation matrices. Choosing the distance metric is highly non-trivial, since it is hard to predict its impact in the final application/study of interest, e.g., \gls{OP} evaluation of slow-\gls{FAMA}, where interference plays the main role. We next propose some sensible heuristic solutions. 

\rev{\subsubsection{Equal-size blocks}

As a natural consequence of the results in Section \ref{sec:SpectralAnalysis}, we can evenly group the fluid antenna ports into $B$ independent blocks of equal size and constant correlation, i.e., we split the fluid antenna aperture into $B$ ``spatial coherence blocks", hence establishing a clear parallelism with temporal block fading. As shown in Appendix \ref{app:EigBlockMatrix}, each block $\mat{A}_b$ in \eqref{eq:SigmaHat} yields the set of eigenvalues $\{\widehat{\rho}_{n'}\}_{n'=1}^{L_b}$ with
\begin{equation}
    \widehat{\rho}_{n'} = \begin{cases}
        (L_b-1)\mu_b^2 + 1 & \text{if } n' = 1\\
        1-\mu_b^2 & \text{if } n' = 2,\dots, L_b
    \end{cases}. \label{eq:eig_blocks}
\end{equation}
Importantly, if $\mu_b\rightarrow 1$, then the multiple eigenvalues at $1-\mu_b^2$ are conveniently close to $0$, and each block will produce a single dominant eigenvalue at $(L_b-1)\mu_b^2 + 1$. Therefore, if we set $\mu_b$ $\forall$ $b$ large, we can view each block as an approximation to  a dominant eigenvalue of $\bm{\Sigma}$. Setting $B$ equal to the number of dominant eigenvalues of $\bm{\Sigma}$, the approximated correlation matrix renders the same \textit{degrees of freedom} than $\bm{\Sigma}$. Moreover, since all the blocks are equal, $L_b = N/B$ $\forall$ $b$ (ignoring rounding errors). Note that this method is equivalent to defining a spatial coherence distance spanning over $N/B$ ports (half-wavelength for, e.g., Jakes's and Clarke's models), and assuming that the ports withing each block are equally correlated with correlation coefficient close to one.}

\rev{\subsubsection{Approximating the spectrum of $\bm{\Sigma}$}

The previous method, albeit consistent with sampling theory and the block fading parallelism, only focuses on the number of spatial blocks and disregards the specific distribution of the eigenvalues of $\bm{\Sigma}$. A further step is aiming to approximate the spectral characteristics of the target correlation matrix. To that end, first notice} that the form of the eigenvectors of $\widehat{\bm{\Sigma}}$ is already imposed by its block-diagonal structure, and these will likely differ from those of the target correlation matrix $\bm{\Sigma}$. A sensible choice is then to focus on the approximation of the spectrum (set of eigenvalues) of $\widehat{\bm{\Sigma}}$. \rev{Using \eqref{eq:eig_blocks} and keeping $\mu_b\rightarrow 1$, we can again define one block per relevant eigenvalue in $\bm{\Sigma}$. In this case, however, we aim to 
%This is motivated by  the previous statistical results in Corollaries \ref{coro:Clarke3D} and \ref{coro:Jakes}, where the number of relevant eigenvalues of $\bm{\Sigma}$ was shown to be proportional to the number of half-wavelengths. }
%is linear with the length (one-dimensional) and area (2D) of the fluid antenna.
fine-tune the block parameters ($L_b$ and $\mu_b$) to approximate the corresponding dominant eigenvalue. Intuitively, this choice \textit{varies} the ``spatial coherence distance" between blocks, while keeping independence between them.} %Intuitively, this choice nicely connects with the ``spatial coherence interval'' concept and the fact that channels (ports) with half-wavelength spacing should be nearly independent, thereby modeled by independent blocks.

% \dmj{Esta eleccion (fijar el mu alto, para tener correspondencia bloque-autovalor dominante) se puede justificar tambien con la conexion con sampling theory. Es decir, hay tantos autovalores dominantes como half-wavelenghts, y tiene sentido (por el tema de 'coherencia' espacial) que haya tantos bloques como half-wavelenghts. Por tanto, queremos tener un autovalor dominante por cada bloque, y para ello hay que hacer mu grande.} \pre{Añadi algo mas al parrafo anterior según lo que comentas aquí ;)}

More specifically, consider the arbitrary small threshold $\rho_\text{th}$, and define the set of sorted eigenvalues of $\bm{\Sigma}$ larger than $\rho_\text{th}$ as $\mathcal{S}(\rho_\text{th}) = \{\rho_n | \rho_n > \rho_\text{th}, n=1,\dots,N\}$. Hence, we define one block in \eqref{eq:SigmaHat} per each eigenvalue in $\mathcal{S}$, i.e., $B = |\mathcal{S}|$. Assume also, for simplicity, that $\mu_b = \mu$ for $b=1,\dots ,B$, with $\mu$ close to $1$. As shown in the previous section, $B\ll N$ when $N$ becomes large. Therefore, for each $\rho_b$ in $\mathcal{S}$, we select the size $L_b$ of the corresponding block such that $\rho_b \approx (L_b-1)\mu^2 +1 = \widehat{\rho}_b$, i.e., we tune the block size to obtain a good approximation of the eigenvalue\footnote{\rev{$L_b$ can be seen as the result of rounding the corresponding eigenvalue $\rho_b$ to the nearest integer.}}, aiming to solve
\begin{equation}
    \mathop {\text{arg min}}\limits_{\;L_1,\dots,L_B} \;\;\mathop \sum_{b=1}^B \|\rho_b - \widehat{\rho}_b\|_2.
\end{equation}
Note that $\sum_{b=1}^B \rho_b \approx N = \sum_{b=1}^B L_b$; this ensures that, ignoring rounding errors due to the integer $L_b$, we should be able to approximate all the eigenvalues in $\mathcal{S}$. Algorithm \ref{alg:BlockApproximation} describes the complete procedure, which iteratively increases the block size until either the corresponding eigenvalue is well approximated or we run out of degrees of freedom ($\sum_{b}L_b = N$).

\RestyleAlgo{ruled}
\begin{algorithm}[t]
    \caption{Computation of $\widehat{\bm{\Sigma}}$}
    \label{alg:BlockApproximation}
    \textbf{Input}: $\mathcal{S}$, $N$, $\mu$\;
    \textbf{Initialize:} $\mathcal{I} = \{1, \dots, B\}$, $L_b = 0$ for $b = 1,\dots,B$\;
    \While{$\sum_{b=1}^B L_b < N$}{
        \ForEach{$b\in\mathcal{I}$}{
            $L_b\leftarrow L_b + 1$\;
            \If{$|(L_b-1)\mu^2 + 1 - \rho_b| \leq |(L_b)\mu^2 + 1 - \rho_b|$}{
                $\mathcal{I}\leftarrow \mathcal{I}\backslash \{b\}$\;
            }
        }
    }
\textbf{Outputs}: $\{L_1,\dots,L_B\}$
\end{algorithm}

Regarding the choice of $\mu$ and $\rho_\text{th}$ \rev{for either of the approximation methods}, no closed-form solution is available. Ideally, $\mu$ should be as close to 1 as possible, but we have observed that too large values of $\mu$ are detrimental when applying the proposed model to \gls{FAMA} analysis. Thus, $\mu^2\in(0.95, 0.99)$ seems to be a good rule-of-thumb. Besides, since the eigenvalues of $\bm{\Sigma}$ decay rapidly, $\rho_\text{th} = 1$ usually yields accurate enough results.

\begin{figure}[t] 
    \centering
  \subfloat[Comparison with Jakes's model.\label{fig:Eig_comparison_128}]{%
       \includegraphics[trim = {1cm 7.5cm 1cm 7.5cm}, clip, width=0.8\columnwidth]{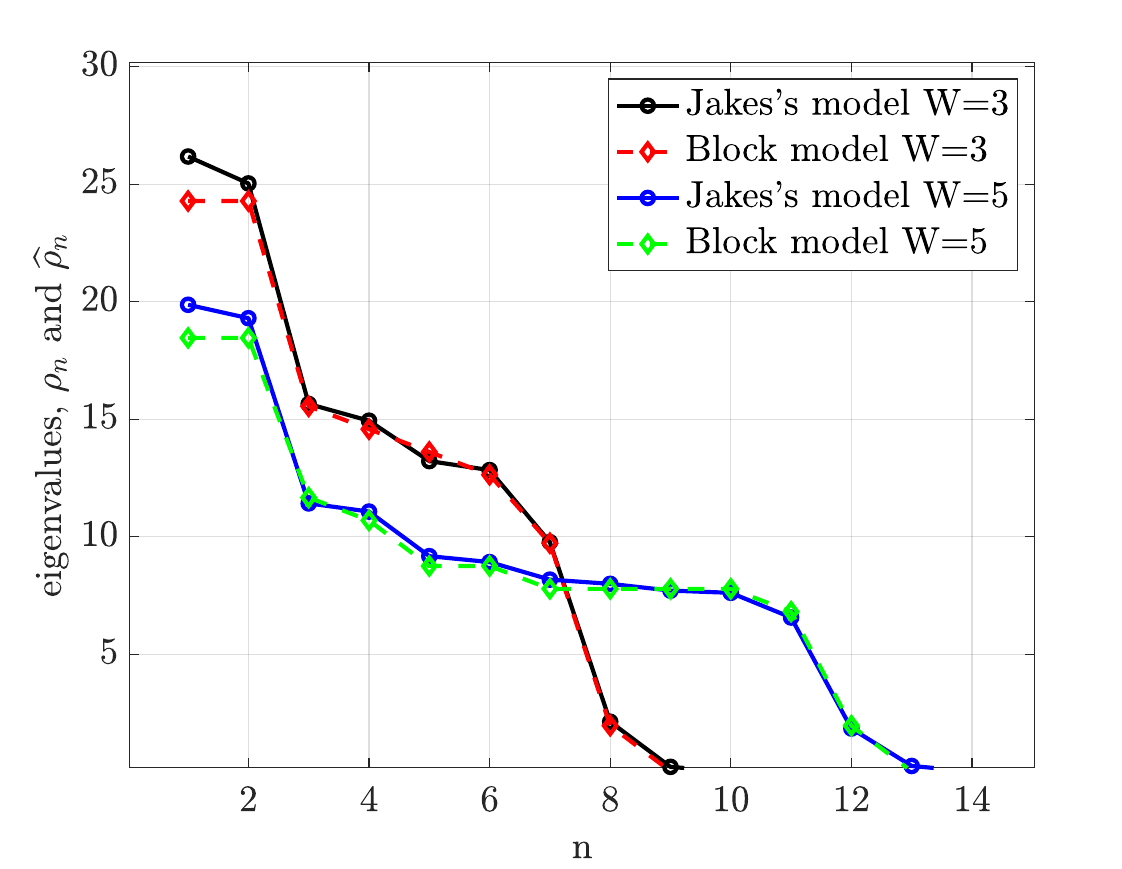}}
    \hfill
  \subfloat[Comparison with 3D Clarke's model.\label{fig:Eig_comparison_512}]{%
        \includegraphics[trim = {1cm 7.5cm 1cm 7.5cm}, clip, width=0.8\columnwidth]{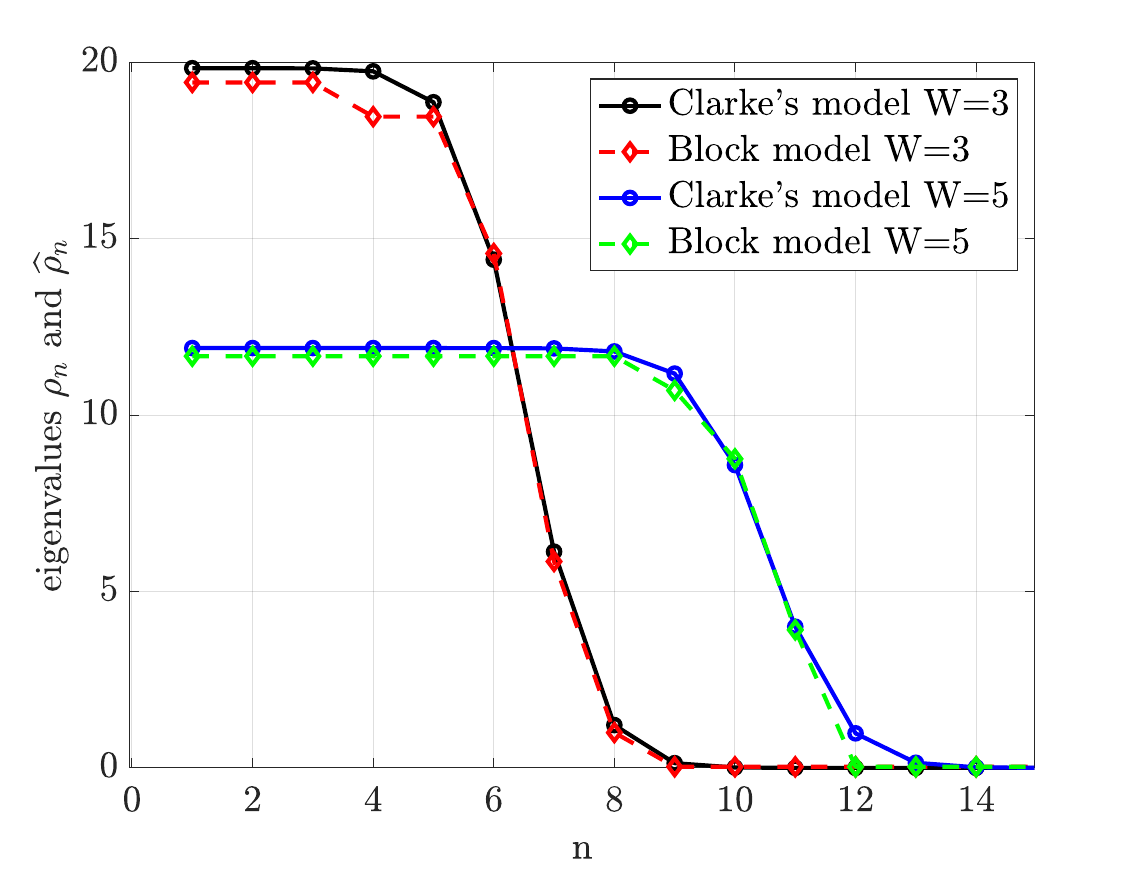}}
  \caption{Comparison of sorted eigenvalues for Jakes's and Clarke's models and block-diagonal correlation matrices according to Algorithm \ref{alg:BlockApproximation}; $N = 120$, $\mu^2 = 0.97$, and $\rho_\text{th} = 1$.}
  \label{fig:Eig_comparison} 
\end{figure}

As an illustrative example, Fig. \ref{fig:Eig_comparison} plots the eigenvalues of both $\bm{\Sigma}$ and $\widehat{\bm{\Sigma}}$ when Algorithm \ref{alg:BlockApproximation} is applied to Jakes's and Clarke's correlation models. Observe that the block-diagonal matrix successfully captures the spectrum of the true correlation matrix. \rev{Note that, in case of setting $\mu_b=\mu$ $\forall$ $b$ in the case of equal block sizes, all the resulting eigenvalues would have the same value.} We shall later evaluate the goodness of the proposed block-diagonal model when used to analyze slow-\gls{FAMA} systems.

\rev{Compared to other classical matrix approximations, such as rank-reduction methods as used in \cite{Khammassi2023, Kiat2023_New}, the block diagonal approach may suffer from a larger approximation error in terms of the correlation matrix itself. However, it considerably eases the mathematical analysis of \gls{FAS}, something that is unfeasible with other methods. Hence, it arises as a trade-off solution between accuracy and tractability. It is also important to note that a tighter approximation to the correlation matrix (lower matrix norm distance) may not directly translate into a better approximation to communication metrics like, e.g., \gls{OP}. 
\begin{remark}
    Albeit the focus here is correlated Rayleigh fading because of its popularity in \gls{FAS} and multi-antenna systems, the block-diagonal approximation can be applied to \textit{any stochastic} channel, such as Rician or the more general model in \cite{Zhu2023Model}.
\end{remark}}

\subsection{Independent antennas equivalent model}

\label{sec:iid_approx}
The proposed block-diagonal correlation aims to approximate one dominant eigenvalue of $\bm{\Sigma}$ per block, and we have seen that $\mu\rightarrow 1$ is needed to that end. A natural further approximation is directly setting $\mu = 1$, i.e., perfect correlation between the ports within each block. In such case, each block can be regarded as a single antenna, and the \gls{FAS} directly approximated by a collection of $B$ independent antennas. As shown in \rev{Lemma \ref{lemma:Clarke3D} and Corollary \ref{coro:Jakes}}, the number of equivalent independent antennas is related to the number of half-wavelengths contained in the aperture. This approximation can be seen as a rough (but extremely simple) bound for performance, helpful to characterize the gain brought by ``oversampling'' in FAS. 

%------------------------------------------------------------------------------------------------------------------
% APPLICATION TO SLOW FAMA
%------------------------------------------------------------------------------------------------------------------
\section{Application to Slow-FAMA}
\label{sec:Slow-FAMA}

\subsection{System model}

Consider the system model of \cite{Wong2022}, where a \gls{BS} with several antennas simultaneously communicates  with $U$ users. The symbols intended for each user are directly transmitted from different \gls{BS} antennas (without precoding) and the users are equipped with a fluid antenna of $N$ ports (either 1D or 2D). Thus, in the downlink, the received signal at the $n$-th port of user $u$'s fluid antenna is
\begin{equation}
    r_n^{(u)} = s_u h_n^{(u,u)} + \sum_{\substack{\tilde{u} = 1 \\ \tilde{u}\neq u} }^U s_{\tilde{u}} h_n^{(\tilde{u},u)} + w_n^{(u)}, \label{eq:SignalModel}
\end{equation}
where $s_u\in\mathbb{C}$ is the symbol intended for user $u$, $h_n^{(\widetilde{u},u)}\in\mathbb{C}$ is the channel gain from the \gls{BS} antenna dedicated for transmitting user $\tilde{u}$'s signal to the $n$-th fluid antenna port, and $w_n^{(u)}\sim\mathcal{CN}(0,\sigma_w^2)$ is the noise term. The symbols $s_u$ for different users are assumed to be independent with $\mathbb{E}[|s_u|^2] = 1$ $\forall$ $u$, and the vector channel $\vec{h}^{\widetilde{u},u} = (h_1^{(\tilde{u},u)} \,\,\, h_2^{(\tilde{u},u)} \,\,\, \dots \,\,\, h_N^{(\tilde{u},u)})$ is jointly Gaussian, i.e., $\vec{h}^{\widetilde{u},u}\sim\mathcal{CN}_N(\vec{0}, \sigma_\alpha^2\bm{\Sigma})$.
%As detailed in Section \ref{subsec:FAS_channel_model},

Assuming block fading and an interference-limited scenario, within each channel coherence time, user $u$ selects the antenna port $\widehat{n}_u$ maximizing the \gls{SIR}, i.e.,
\begin{equation}
    \widehat{n}_u = \mathop{\text{arg max}}_n \frac{\left|h_n^{(u,u)}\right|^2}{\sum_{\substack{\tilde{u} = 1 \\ \tilde{u}\neq u} }^U \left| h_n^{(\tilde{u},u)}\right|^2}.
    \label{eq:SIR_Rayleigh}
\end{equation}
\rev{The port selection in \eqref{eq:SIR_Rayleigh} requires perfect channel and interference knowledge at each port of the fluid antenna, which in practice may imply a large overhead---the reader is referred to, e.g., \cite{Wang2023, Chai2022} for further discussions on practical port selection methods. However, \eqref{eq:SIR_Rayleigh} represents the performance limit of \gls{FAMA}, and helps to answer the fundamental questions posed throughout this paper. }

\subsection{Outage probability analysis}
\label{subsec:Outage}

Defining by $\gamma$ the minimum acceptable \gls{SIR} for user $u$, we are interested in the probability 
\begin{equation}
    P_\text{out}(\gamma) = P\left(\max_n\frac{\left|h_n^{(u,u)}\right|^2}{\sum_{\substack{\tilde{u} = 1 \\ \tilde{u}\neq u} }^U \left| h_n^{(\tilde{u},u)}\right|^2} < \gamma \right), \label{eq:Pout_true}  
\end{equation}
which, under the correlation models in \eqref{eq:3DClarke} and \eqref{eq:Jakes}---or, in general, under an arbitrary spatial correlation function given by \eqref{eq:Sigma_General}---is very complex, if not impossible \cite{Psomas2023, Khammassi2023, Kiat2023, Kiat2023_New}. Thus, we propose approximating the \gls{OP} by replacing the true correlation matrix $\bm{\Sigma}$ of the channel by our block-correlation matrix $\widehat{\bm{\Sigma}}$ (Algorithm \ref{alg:BlockApproximation}), i.e., we assume $\vec{h}^{\widetilde{u},u}\sim\mathcal{CN}_N(\vec{0}, \sigma_\alpha^2\widehat{\bm{\Sigma}})$. Note that this approximation is valid for both 1D and 2D fluid antennas and, in principle, for any correlation matrix beyond the examples given in this paper. The OP is approximated by
\begin{equation}
	P_\text{out}(\gamma) \approx \widehat{P}_\text{out}(\gamma)= P\left(\max_n \frac{X_n}{Y_n} < \gamma \right), \label{eq:ObjFun}
\end{equation}
where the variables $X_n$ and $Y_n$ are block-wise defined as
\begin{align}
	X_n =& \left(x_n^{(u,u)} + \frac{\mu}{\sqrt{1-\mu^2}}x_{b(n)}^{(u,u)}\right)^2	\notag \\
     &+ \left(y_n^{(u,u)} + \frac{\mu}{\sqrt{1-\mu^2}}y_{b(n)}^{(u,u)}\right)^2,\label{eq:Xn} \\
%\end{align}
%\begin{align}    
	Y_n =& \sum_{\widetilde{u}=1, \widetilde{u} \neq u}^U\left(x_n^{( \widetilde{u} ,u)} + \frac{\mu}{\sqrt{1-\mu^2}}x_{b(n)}^{( \widetilde{u} ,u)}\right)^2 \notag \\
    &+ \left(y_n^{( \widetilde{u} ,u)} + \frac{\mu}{\sqrt{1-\mu^2}}y_{b(n)}^{( \widetilde{u} ,u)}\right)^2.
\end{align}
All the random variables above (namely, $x_j^{(u,u)}, y_j^{(u,u)}, x_j^{(\widetilde{u},u)}$ and $y_j^{(\widetilde{u},u)}$ for $j=n,b$) are independent Gaussian variables with zero mean and unit variance. Besides, $b(n)$ is the block index, i.e., $b(n) = 1$ for $n = 1,\dots, L_1$, $b(n) = 2$ for $n = L_1+1,\dots, L_2$, and so on. For ease of notation, the dependence with $n$ is dropped in the following.
%The resulting expression for the outage probability is given next.

\begin{lemma}
    The \gls{OP} of slow-\gls{FAMA}
    %in \eqref{eq:Pout_true} 
    is approximated by
    \begin{align}
	\widehat{P}_\text{out}(\gamma) =& \prod_{b=1}^B \int_0^\infty \int_0^\infty \frac{\widetilde{r}_b^{U-2}e^{-\frac{r_b + \widetilde{r}_b}{2}}}{2^{U}\Gamma(U-1)}\left[G(\gamma ; r_b, \widetilde{r}_b)\right]^{L_b}\dif r_b \dif \widetilde{r}_b, \label{eq:Pout} 
\end{align}
with $G(\gamma ; r_b, \widetilde{r}_b)$ given by \eqref{eq:G} at the top of next page, where $Q_p(\cdot,\cdot)$ is the $p$-th order Marcum-Q function and $I_\nu(\cdot)$ is the $\nu$-th order modified Bessel function of the first kind. \label{lemma:Pout}
\end{lemma}
\begin{IEEEproof}
    See Appendix \ref{app:Outage}.
\end{IEEEproof}

\begin{figure*}[t]
\begin{align}
G(\gamma ; r_b, \widetilde{r}_b) =&\; Q_{U-1}\left(\sqrt{\frac{\mu^2\gamma \widetilde{r}_b}{(1-\mu^2)(\gamma+1)}}, \sqrt{\frac{\mu^2 r_b}{(1-\mu^2)(\gamma+1)}}\right)- \left(\frac{1}{\gamma+1}\right)^{U-1} \exp{-\frac{\mu^2}{2(1-\mu^2)}\frac{\gamma\widetilde{r}_b+r_b}{\gamma+1}}\notag \\
	& \times\sum_{k=0}^{U-2}\sum_{j=0}^{U-k-2}\frac{(U-(j+k)-1)_j}{j!}\left(\frac{r_b}{\widetilde{r}_b}\right)^{\frac{j+k}{2}} (\gamma+1)^k\gamma^{\frac{j-k}{2}}I_{j+k}\left(\frac{\mu^2\sqrt{\gamma r_b\widetilde{r}_b}}{(1-\mu^2)(\gamma+1)}\right). \label{eq:G}
\end{align}
\hrulefill
\vspace*{4pt}
\end{figure*}

Note that each product term in \eqref{eq:Pout}, representing the contribution of each block in \eqref{eq:SigmaHat}, has exactly the same form as \cite[Eq. (21)]{Wong2022}; this guarantees the same analytical tractability, as compared with the simple constant correlation model of \cite{Wong2022}. To alleviate the computational effort of evaluating several double-integrals, Gauss-Laguerre quadrature can be used, leading to
\begin{equation}
    \widehat{P}_\text{out}(\gamma) \approx \prod_{b=1}^B \tfrac{1}{\Gamma(U-1)}\sum_{m=1}^M \sum_{\widetilde{m}=1}^{\widetilde{M}} w_m \widetilde{w}_{\widetilde{m}} \left[G(\gamma ; 2 x_m, 2\widetilde{x}_{\widetilde{m}})\right]^{L_b}, \label{eq:Pout_Quadrature}
\end{equation}
where $x_m$ for $m = 1,\dots, M$ and $\widetilde{x}_{\widetilde{m}}$ for $\widetilde{m} = 1,\dots, \widetilde{M}$ are, respectively, the roots of the Laguerre polynomials $L_M(x)$ and $L_M^{U-2}(x)$, and
\begin{align}
    w_m &= \frac{x_m}{(M+1)^2\left[L_{M+1}(x_m)\right]}, \\
    \widetilde{w}_{\widetilde{m}} &= \frac{\Gamma(\widetilde{M}+U-1)\widetilde{x}_{\widetilde{m}}}{\widetilde{M}!(\widetilde{M}+1)^2\left[L^{U-2}_{\widetilde{M}+1}(\widetilde{x}_{\widetilde{m}})\right]}.
\end{align}
Equation \eqref{eq:Pout_Quadrature} considerably reduces the computational effort, since the function $G(\cdot)$ needs to be computed only once for each pair $(x_m, \widetilde{x}_{\widetilde{m}})$, regardless of the number of blocks $B$.

Interestingly, a further approximation to \eqref{eq:Pout} can be obtained by particularizing for $\mu\rightarrow 1$; recall that our proposed block-diagonal approximation requires large (close to 1) values of $\mu$.  %in Section \ref{sec:Blockapprox}.
First notice that the sum in \eqref{eq:G} is dominated by the exponential term as $\mu\rightarrow 1$ and can thus be neglected, i.e., 
\begin{align}
    [G(\gamma; &r_b, \widetilde{r}_b)]^{L_b} \approx \notag \\
    &\left[Q_{U-1}\left(\sqrt{\tfrac{\mu^2\gamma \widetilde{r}_b}{(1-\mu^2)(\gamma+1)}}, \sqrt{\tfrac{\mu^2 r_b}{(1-\mu^2)(\gamma+1)}}\right)\right]^{L_b}. \label{eq:G_approx}
\end{align}
Besides, the spectral analysis in Section \ref{sec:SpectralAnalysis} highlights that, for very large $N$, only a very small fraction of eigenvalues is relevant. This implies that $B \ll N$, and since $\sum_{b}^B L_b = N$, we can expect very large values for each $L_b$ (at least for most of the dominant eigenvalues). A large exponent strengthens the sigmoid behaviour of $Q_p(\cdot)$, and in the limit ($L_b\rightarrow \infty$) $(Q_p(\cdot))^{L_b}$ becomes a simple step function. We propose to introduce this approximation in \eqref{eq:Pout}, as stated next.
\begin{corollary}
    \label{coro:ApproxOutage}
    For $\mu\rightarrow 1$, the approximation to the \gls{OP} of slow-\gls{FAMA} in \eqref{eq:Pout} is further approximated by
    \begin{align}
        \widehat{P}_\text{out}(\gamma) \approx \prod_{b=1}^B \left[1 - \frac{2^{1-U}}{\Gamma(U-1)}\int_0^\infty \widetilde{r}_b^{U-2} e^{(-\widetilde{r}_b - \delta(\widetilde{r}_b))/2}\dif \widetilde{r}_b \right], \label{eq:Pout_simplified}
    \end{align}
where $\delta(\widetilde{r}_b)$ is given by
\begin{equation}
    \label{eq:threshold}
    \delta(\widetilde{r}_b) = \left(\sqrt{\gamma \widetilde{r}_b} + \frac{\frac{(U-\frac{3}{2})(1+\gamma)^{\frac{1}{2}}}{\mu(1-\mu^2)^{-\frac{1}{2}}} - \frac{L-1}{\sqrt{2\pi}}\sqrt{\gamma\widetilde{r}_b}}{\frac{(L-1)(U-\frac{3}{2})}{\sqrt{2\pi}} + \sqrt{\frac{\mu^2\gamma\widetilde{r}_b}{(1-\mu^2)(1+\gamma)}}}\right)^2.
\end{equation}
\end{corollary}
\begin{IEEEproof}
    See Appendix \ref{app:ApproxOutage}.
\end{IEEEproof}
Once again, a computationally-efficient formula is obtained by applying Gauss-Laguerre quadrature to \eqref{eq:Pout_simplified}, giving
\begin{equation}
    \label{eq:Pout_Quadrature_simplified}
    \widehat{P}_\text{out}(\gamma) \approx \prod_{b=1}^B \left[1 - \frac{1}{\Gamma(U-1)}\sum_{\widetilde{m}}^{\widetilde{M}}\widetilde{w}_{\widetilde{m}} e^{-\frac{1}{2}\delta(2\widetilde{x}_{\widetilde{m}})}\right],
\end{equation}
where $\widetilde{w}_{\widetilde{m}}$ and $\widetilde{x}_{\widetilde{m}}$ are as in \eqref{eq:Pout_Quadrature}.

In Corollary \ref{coro:ApproxOutage}, $\delta(\widetilde{r}_b)$ is considered as a threshold such that $[G(\gamma; r_b, \widetilde{r}_b)]^{L_b} = 1$ for ${r}_b < \delta(\widetilde{r}_b)$ and $[G(\gamma; r_b, \widetilde{r}_b)]^{L_b} = 0$ for ${r}_b > \delta(\widetilde{r}_b)$, so that the step function approximation mentioned previously is applied. The specific value in \eqref{eq:threshold} is chosen such that $[G(\gamma; \delta(\widetilde{r}_b), \widetilde{r}_b)]^{L_b} \approx \frac{1}{2}$, corresponding to the point of steepest slope. Interestingly, other choices of $\delta(\widetilde{r}_b)$ are possible, yielding different upper and lower bounds of the \gls{OP}.
%Corollary \ref{coro:ApproxOutage} provides a much simpler expression for the OP, with the accuracy of the approximation given, to some extent, by the choice of $\delta(\widetilde{r}_b)$. The threshold in \eqref{eq:threshold} is chosen such that $[G(\gamma; r_b, \widetilde{r}_b)]^{L_b} \approx \frac{1}{2}$, corresponding to the point of steepest slope. Interestingly enough, \eqref{eq:Pout_simplified} allows to arbitrarily choose $\delta(\widetilde{r}_b)$ so that different upper and lower bounds are obtained. 
For instance, for an arbitrary Marcum's function $Q_p(\alpha, z)$, it is known that $Q_p(\infty, z) = 1$ while $Q_p(\alpha, \infty) = 0$. When both $\alpha$ and $z$ are very large, as in \eqref{eq:G_approx} due to the term $1-\mu$ in the denominators, the trivial choice $z = \alpha$ is motivated to derive the threshold $\delta(\widetilde{r}_b)$. Comparing with \eqref{eq:G_approx}, this yields $\delta(\widetilde{r}_b) = \gamma\widetilde{r}_b$, which introduced in \eqref{eq:Pout_simplified} leads to
\begin{equation}
    \widehat{P}_\text{out}^\text{iid}(\gamma) \approx \left(1 - \frac{1}{(\gamma+1)^{U-1}}\right)^B, \label{eq:Pout_ind}
\end{equation}
where \cite[Eq. (3.351 3)]{Gradshteyn2007} has been used. Remarkably, the above expression is exactly the \gls{OP} obtained by $B$ independent antennas, as proved in Appendix \ref{app:OutageIID}. Hence, the block-diagonal correlation model analytically connects with the equivalent model for independent antennas discussed in Section \ref{sec:iid_approx}. Relating \eqref{eq:Pout_ind} to \eqref{eq:Pout_simplified}, we can approximately obtain the gain (in terms of \gls{OP}) of slow-\gls{FAMA} over a conventional array with $B$ independent antennas (with $\lambda/2$ spacing), i.e., the gain rendered by the oversampling of the fluid antenna, with ports densely packed within each $\lambda/2$ spacing. Specifically, letting $\Delta P_\text{out}^b$ be the difference in \gls{OP} achieved by the $b$-th block in \eqref{eq:Pout_ind} and \eqref{eq:Pout_simplified}, i.e., the difference in \gls{OP} between a single antenna and a block of correlated elements, we have
\begin{align}
 \Delta P_\text{out}^b =&  \tfrac{2^{1-U}}{\Gamma(U-1)}\int_0^\infty \widetilde{r}_b^{U-2} e^{(-\widetilde{r}_b - \delta(\widetilde{r}_b))/2}\dif \widetilde{r}_b -\frac{1}{(\gamma+1)^{U-1}}, \label{eq:DeltaPout}
\end{align}
and, therefore,
\begin{equation}
    \widehat{P}_\text{out}(\gamma) = \prod_{b=1}^B \left[1 - \frac{1}{(\gamma+1)^{U-1}} - \Delta P_\text{out}^b\right].
\end{equation}

\begin{figure}[t]
    \centering
    % Trim option due to figure saved directly from Matlab
    \includegraphics[trim = {0 7.5cm 0 7.5cm}, clip, width = \columnwidth]{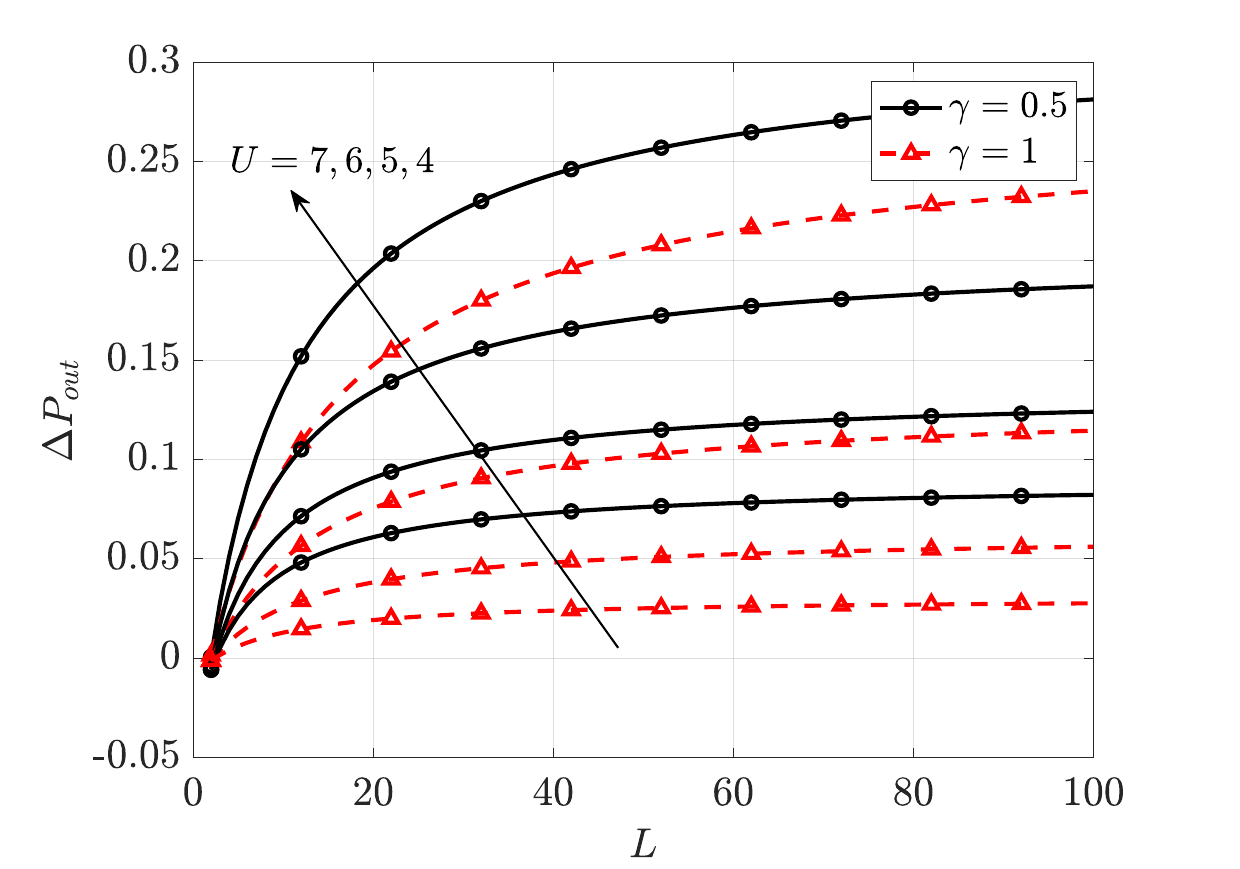}
    \caption{Gain, in terms of outage probability, achieved by each block of correlated elements ($\mu^2 = 0.97$) in \eqref{eq:Pout_simplified} w.r.t. a single antenna element.}
    \label{fig:DeltaPout}
\end{figure}

As an example, the gain $\Delta P_\text{out}^b$ for an arbitrary block is depicted in Fig. \ref{fig:DeltaPout} for different numbers of users $U$, block sizes $L$ and thresholds $\gamma$. Interestingly, we observe the expected saturation effect as $L$ increases, since a large block size implies a large ratio $N/W$ and, hence, a dense fluid antenna. Besides, the gain rendered by the \gls{FAS} is less noticeable as the number of users increases, due to the higher interference, and as the threshold $\gamma$ is increased. Note also that this is a per-block gain, which stacks with the gains of the other blocks in \eqref{eq:Pout_simplified}, meaning that a \gls{FAS} with more blocks will deliver higher returns. Recall also that, according to our proposed model, a block approximately represents a spatial coherence interval of half-wavelength. Higher returns are thus expected in \gls{FAS} with larger apertures (accommodating more blocks).  

\rev{\subsection{Special case: Single user FAS}

A particular case of \gls{FAS} is when only one user is served---i.e., a single antenna at the \gls{BS} and a user equipped with a fluid antenna. Albeit out of the \gls{FAMA} context (since there is no multiple access), this case may be of interest to gain insight in the saturation of the fluid antenna. When considering a single user, the port selection in \eqref{eq:SIR_Rayleigh} simplifies to
\begin{equation}
    \widehat{n}_u = \mathop{\text{arg max}}_n \left|h_n^{(u,u)}\right|^2,
\end{equation}
and, therefore, the \gls{OP} can be approximated by
\begin{equation}
     \left.\widehat{P}_\text{out}(\gamma)\right|_{U=1} = P\left(\max_n \left\{(1-\mu^2)X_n\right\} < \gamma\right),
\end{equation}
with $X_n$ as in \eqref{eq:Xn}\footnote{Note that the scaling factor $(1-\mu^2)$ is needed to preserve the correlation structure and the power of the channel realization at each port}. As there is no interference, $\widehat{P}_\text{out}(\gamma)$ is thus readily obtained from \eqref{eq:JointCDF_Xk}, leading to
\begin{align}
    \left.\widehat{P}_\text{out}(\gamma)\right| =& \prod_{b=1}^B\int_0^\infty\frac{1}{2}e^{-r_b/2}\notag \\
    &\times\left[1-Q_1\left(\sqrt{\frac{\mu^2r_b}{1-\mu^2}},\sqrt{\frac{\gamma}{1-\mu^2}}\right)\right]^{L_b} \text{d}r_b.
\end{align}

Notice that, since there is no interference, in this case $\gamma$ represents the \gls{SNR} (and not the \gls{SIR}) threshold.}
%------------------------------------------------------------------------------------------------------------------
% NUMERICAL RESULTS
%------------------------------------------------------------------------------------------------------------------
\section{Performance evaluation of slow-FAMA}
\label{sec:Numerical}

In this section, we validate the proposed spatial correlation model, and assess the performance of slow-\gls{FAMA} systems. Specifically, we are interested in \textit{i)} the impact of the fluid antenna size, \textit{ii)} the saturation effect as the number of ports increases, \textit{iii)} the number of simultaneous users supported by the system, and \textit{iv)} whether the derived approximations for the \gls{OP} can capture all these effects.

\subsection{One-dimensional FAS}

Consider a linear fluid antenna along the $x$ axis of length $W$ (normalized by the wavelength) and $N$ equally spaced ports. The proposed approximations for the \gls{OP} in slow-\gls{FAMA} \eqref{eq:Pout} and \eqref{eq:Pout_simplified}, under the proposed block-correlation model (obtained from Algorithm \ref{alg:BlockApproximation}), are evaluated with the quadrature expressions in \eqref{eq:Pout_Quadrature} and \eqref{eq:Pout_Quadrature_simplified}, with $M=\widetilde{M} = 30$.

\begin{figure}[t]
    \centering
    \includegraphics[ width = 0.9\columnwidth]{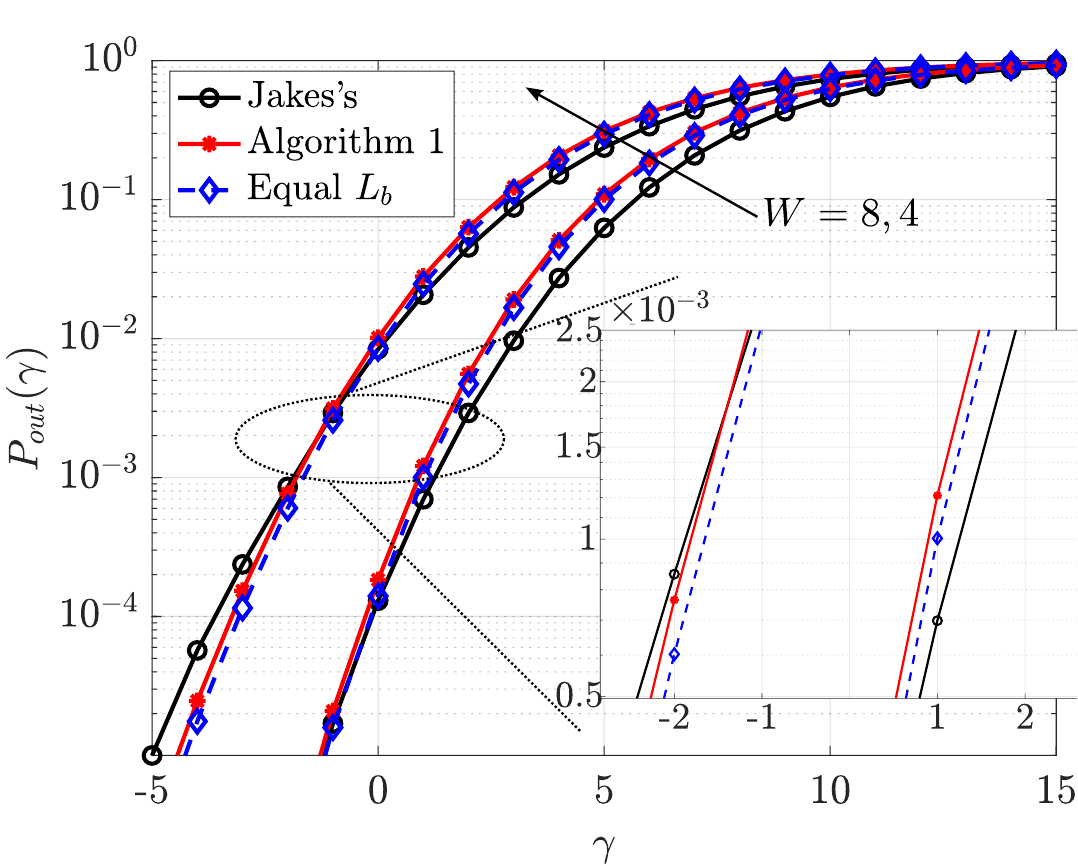}
    \caption{\rev{OP for different SINR thresholds and antenna lengths; Jakes's model \eqref{eq:Sigma_Jakes} is compared with the proposed block-diagonal approximation and the two methods to choose the block sizes (both are evaluated through \eqref{eq:Pout_Quadrature}). $15$ ports per wavelength, $U = 3$, $\mu^2 = 0.96, \rho_\text{th} = 1$.}}
    \label{fig:GammaEvo_Lbequal}
\end{figure}

\begin{figure}[t]
    \centering
    % Trim option due to figure saved directly from Matlab
    \includegraphics[trim = {0 7.5cm 0 7.5cm}, clip, width = \columnwidth]{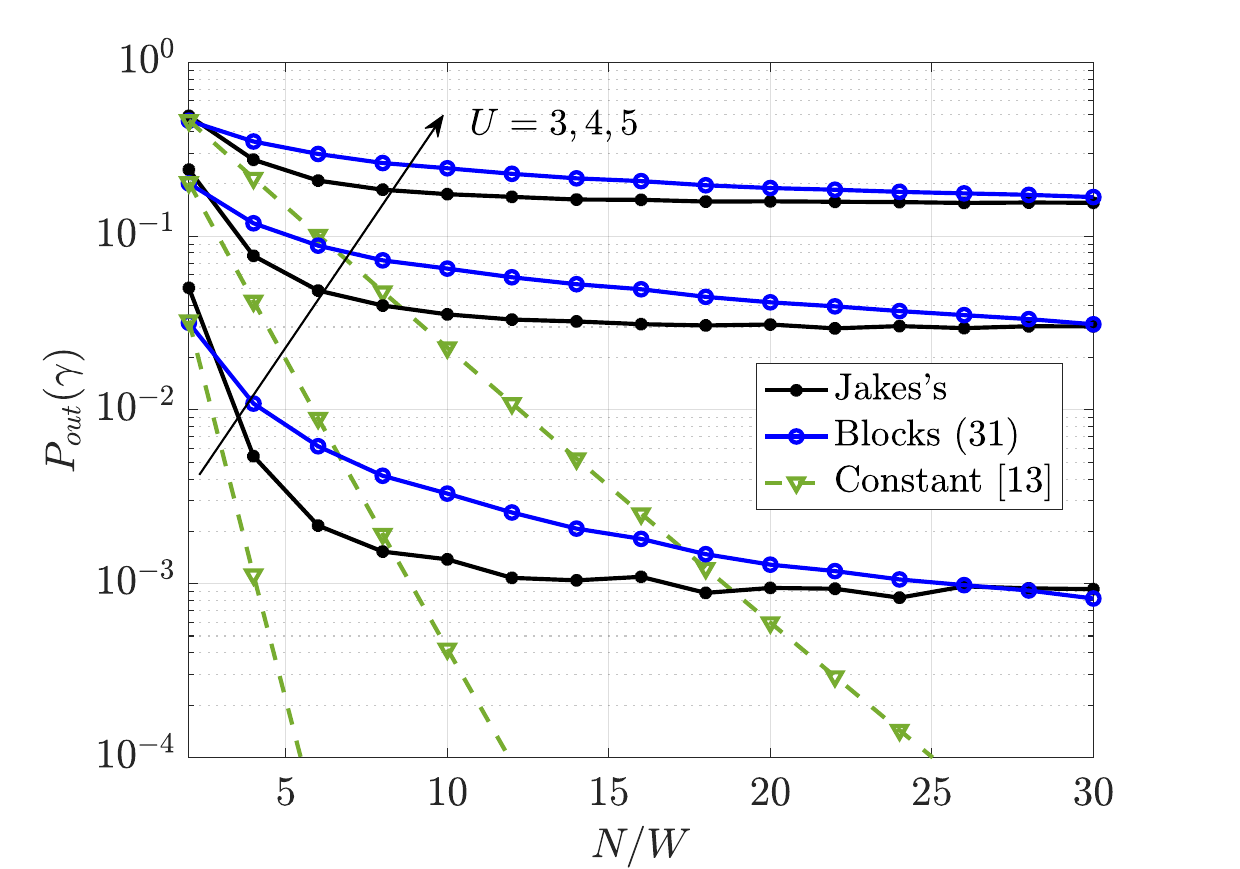}
    \caption{Evolution of outage probability as the fluid antenna is densified; Jakes's model \eqref{eq:Sigma_Jakes} is compared with the proposed approximation and the model in \cite{Wong2022_EL}; $W = 6$, $\gamma = 1$, $\mu^2 = 0.97, \rho_\text{th} = N/100$.}
    \label{fig:Nevo_1D_Jakes}
\end{figure}

\rev{We first compare in Fig. \ref{fig:GammaEvo_Lbequal} the two methods given in Section \ref{sec:Blockapprox} to select the block sizes $L_b$, namely $L_b = N/B$ $\forall$ $b$ (equal block sizes) and Algorithm \ref{alg:BlockApproximation} (which fine-tunes the block sizes to approximate the eigenvalues of the target correlation matrix). As discussed previously, it is hard to predict the impact of the approximation error in terms of matrix distance on the \gls{OP}. In fact, we observe in Fig. \ref{fig:GammaEvo_Lbequal} that both methods yield quite similar results, with only minor differences. However, we should be careful when extrapolating this to other scenarios and metrics beyond \gls{OP} of \gls{SIR}, because similar results for \gls{OP} may not mean similar results elsewhere. Nevertheless, since we focus here on \gls{OP} in \gls{FAMA}, and due to the similarity of the results, we stick to Algorithm \ref{alg:BlockApproximation} in the following results.}

\begin{figure}[t]
    \centering
    % Trim option due to figure saved directly from Matlab
    \includegraphics[trim = {0 7.5cm 0 7.5cm}, clip, width = \columnwidth]{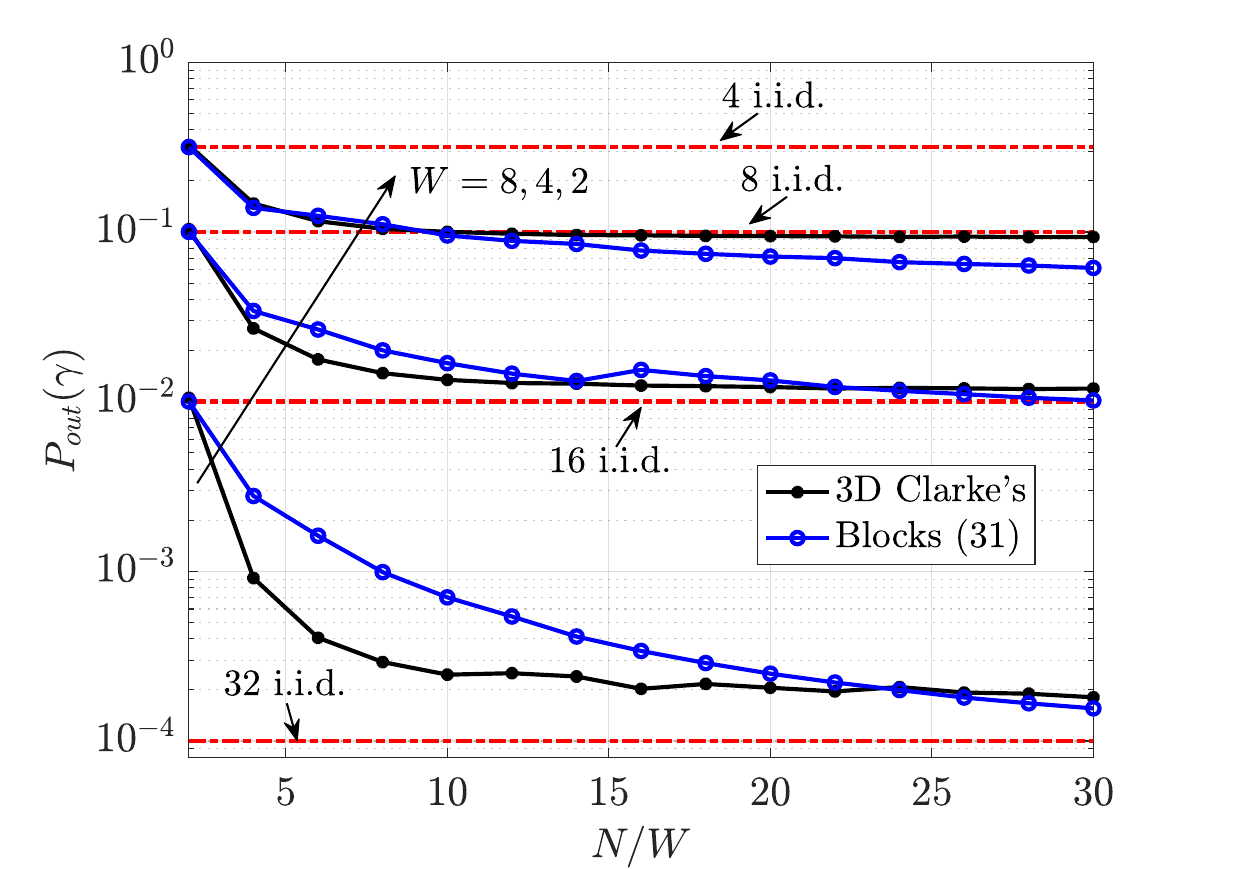}
    \caption{Evolution of outage probability as the fluid antenna is densified; 3D Clarke's model \eqref{eq:3DClarke} is compared with the proposed approximation; $U = 3$, $\gamma = 1$, $\mu^2 = 0.97, \rho_\text{th} = N/100$.}
    \label{fig:Nevo_1D_Clarke}
\end{figure}

\rev{Moving on to the performance of \gls{FAMA} itself, we next show the \gls{OP} achieved by slow-\gls{FAMA}} as $N$ increases for different numbers of users (see Fig. \ref{fig:Nevo_1D_Jakes}). Results under Jakes's model, with correlation structure in \eqref{eq:Sigma_Jakes}, and under the constant correlation model in \cite{Wong2022_EL} are also plotted for comparison. As expected, for fixed $W$, densifying the fluid aperture yields diminishing returns, saturating the system; this effect is also reported in \cite{Khammassi2023, Zhu2023}. Specifically, more than $15$ ports per wavelength ($N/W = 15$) seems to provide no additional benefit due to the already large correlation between adjacent ports. Relating this to Section \ref{sec:SpectralAnalysis}, the resulting Jakes's correlation matrix no longer increases its rank significantly from this point onwards. Interestingly, we observe that the proposed block-diagonal approximation can capture this saturation, while the constant correlation model considerably deviates from the real performance (Jakes's correlation simulation). The same trend is observed for Clarke's model in \eqref{eq:Sigma_Clarke3D} and for different fluid antenna sizes, as illustrated in Fig. \ref{fig:Nevo_1D_Clarke}. This figure also depicts the \gls{OP} achieved by a collection of independent antennas, given by \eqref{eq:Pout_ind}. Note that, given a linear fluid antenna of length $W$, only $2W$ independent antennas (equivalently, $N/W = 2$) could be embedded in the physical aperture, according to the $\lambda/2$ spacing rule-of-thumb. The gain brought by the additional spatial diversity of \gls{FAS} quickly arises with just a few extra ports per wavelength. We also observe that, the larger the fluid antenna, the more the gain w.r.t. the independent antenna system, as predicted by the analysis in Section \ref{subsec:Outage}.
%stacking of more block gains in \eqref{eq:DeltaPout}.
It can also be seen (in both Figs. \ref{fig:Nevo_1D_Jakes} and \ref{fig:Nevo_1D_Clarke}) that the accuracy of the proposed model increases as the fluid antenna is densified. This is consistent with the fact that our block-diagonal approximation is based on asymptotic statistical results for large $N$.
%in Section \ref{sec:SpectralAnalysis}, 
%eing therefore a reasonable solution to analyze the limit performance of \gls{FAS}. 

\begin{figure}[t]
    \centering
    % Trim option due to figure saved directly from Matlab
    \includegraphics[trim = {0 7.5cm 0 7.5cm}, clip, width = \columnwidth]{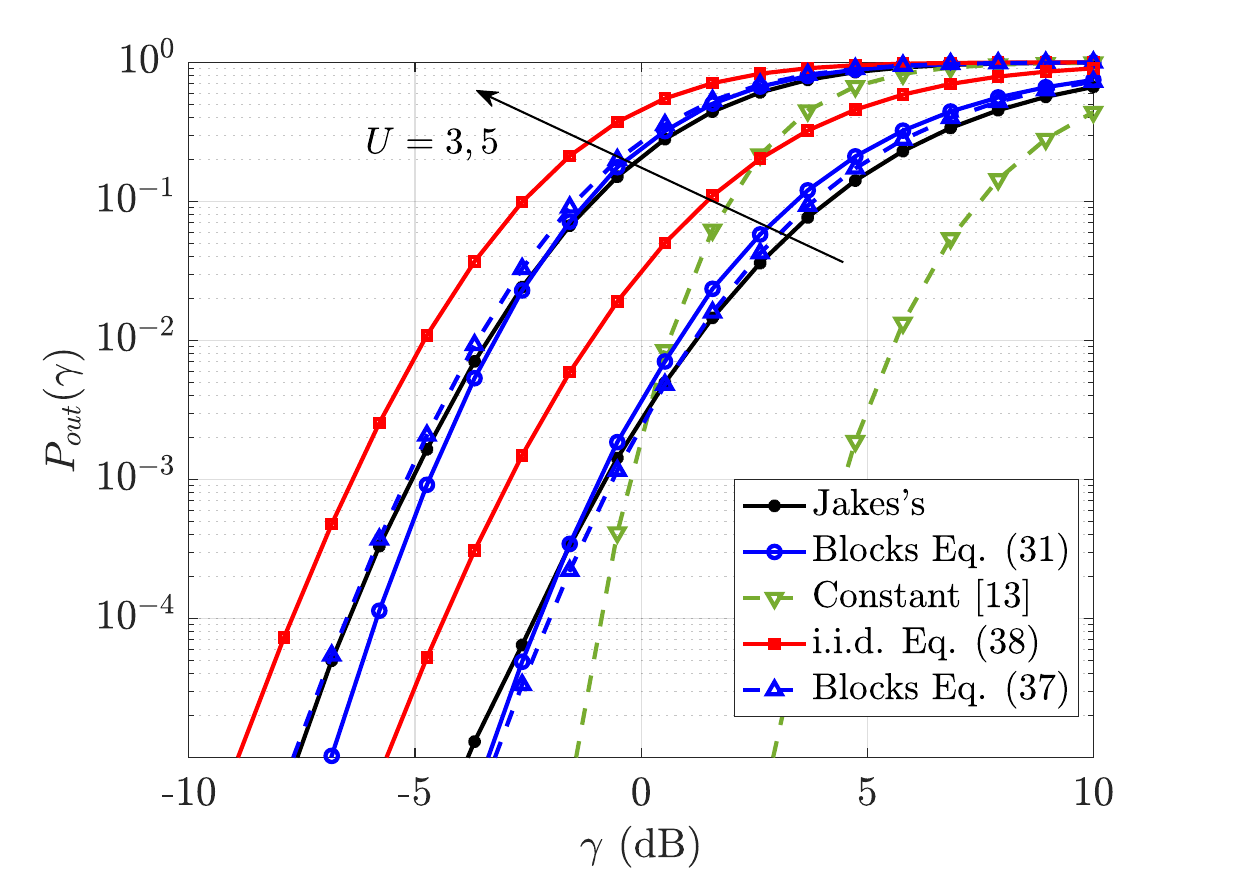}
    \caption{Outage probability vs. $\gamma$; Jakes's model is compared with the proposed approximation, the i.i.d. upper bound, and the constant correlation model in \cite{Wong2022_EL}; $N = 100$, $W = 5$, $\mu^2 = 0.97, \rho_\text{th} = N/100$.}
    \label{fig:Gamma_Evo_Jakes_1D}
\end{figure}

Looking at the left tail of $P_\text{out}(\gamma)$, Fig. \ref{fig:Gamma_Evo_Jakes_1D} depicts the \gls{OP} under Jakes's correlation for different numbers of users. On top of the ground-truth simulated curve, it also shows the one obtained with the block-diagonal approximation \eqref{eq:Pout_Quadrature}, the approximated \gls{OP} expression for large $\mu$ \eqref{eq:Pout_Quadrature_simplified} with the threshold in \eqref{eq:threshold}, the upper bound (independent antennas) in \eqref{eq:Pout_ind} and the result achieved by the constant correlation model in \cite{Wong2022_EL}. As before, we observe that the constant correlation model is overly too optimistic, while both approximated expressions arising from the block-diagonal framework are tight. The upper bound based on independent antennas is naturally pessimistic;
%, since it assumes perfect correlation within each block; 
its usefulness is though given by its simplicity and its ability to somehow capture the general trend of the system (in contrast to the constant correlation model).

\subsection{Two-dimensional FAS}

Consider now a planar fluid antenna of area $W = W_x \times W_z$ contained in the $xz$-plane, where both $W_x$ and $W_z$ are normalized by the wavelength. The total number of ports is $N = N_x\times N_z$, being again equally spaced along both axes. Since Jakes's correlation is not valid for these planar apertures, we here stick to 3D Clarke's model \eqref{eq:Sigma_Clarke3D}. 

\begin{figure}[t]
    \centering
    % Trim option due to figure saved directly from Matlab
    \includegraphics[trim = {0 7.5cm 0 7.5cm}, clip, width = \columnwidth]{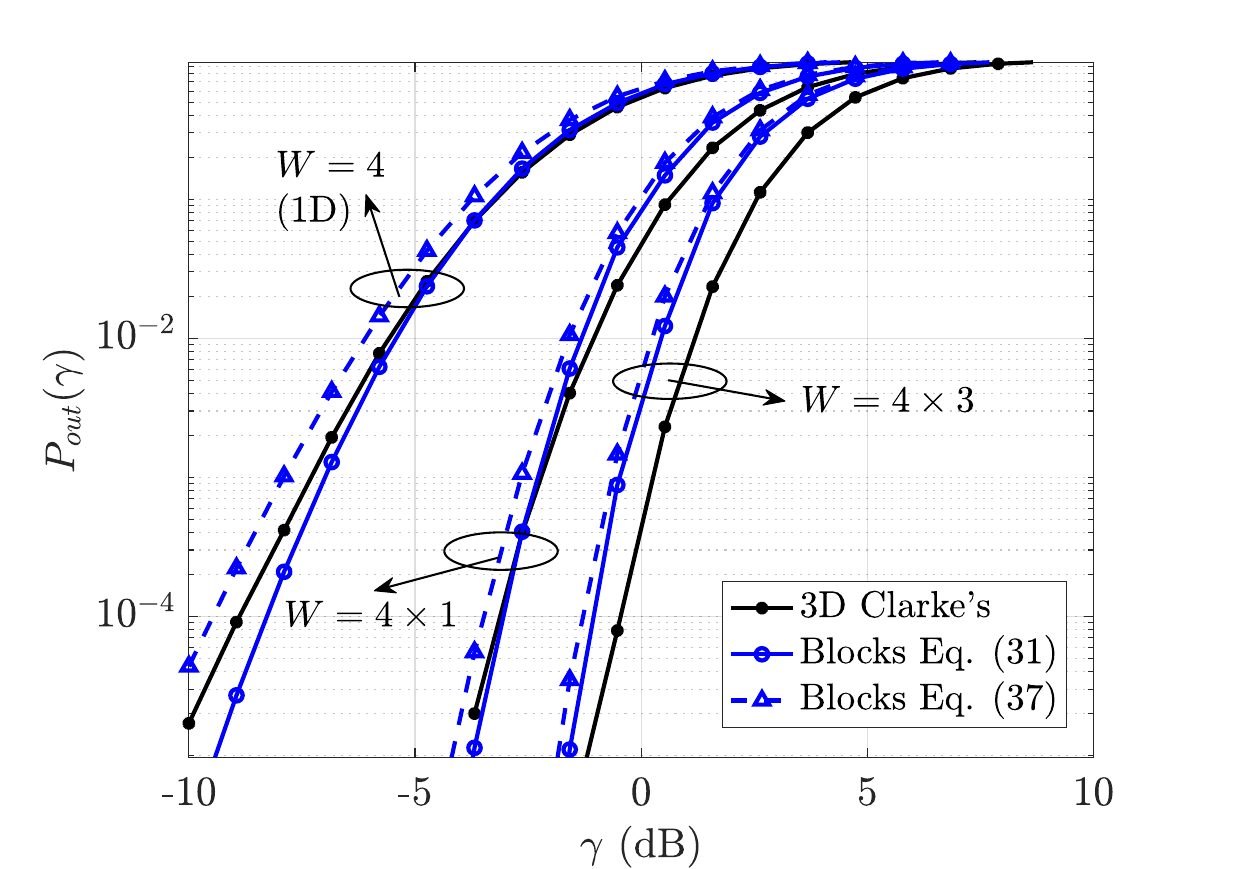}
    \caption{Outage probability vs. $\gamma$ for different fluid antenna sizes; 3D Clarke's \eqref{eq:3DClarke} model is compared with the proposed approximation; $15$ ports per wavelength, $U = 6$, $\mu^2 = 0.97, \rho_\text{th} = 1$.}
    \label{fig:Gamma_Evos_2D}
\end{figure}

Motivated by the promising results reported in \cite{Kiat2023} on 2D \gls{FAS}, we first evaluate in Fig. \ref{fig:Gamma_Evos_2D} the gain (in terms of \gls{OP}) brought by going from 1D to 2D fluid antennas. Interestingly, the gap between a linear antenna of size $W=4$ and a 2D antenna of size $W = 4\times 1$ is very substantial, suggesting that even a thin (albeit planar) surface may be way more beneficial than large linear apertures. On the other hand, the proposed \gls{OP} approximations (both \eqref{eq:Pout} and \eqref{eq:Pout_simplified}) are tightly accurate regardless of the fluid antenna size.

\begin{figure}[t]
    \centering
    % Trim option due to figure saved directly from Matlab
    \includegraphics[trim = {0 7.5cm 0 7.5cm}, clip, width = \columnwidth]{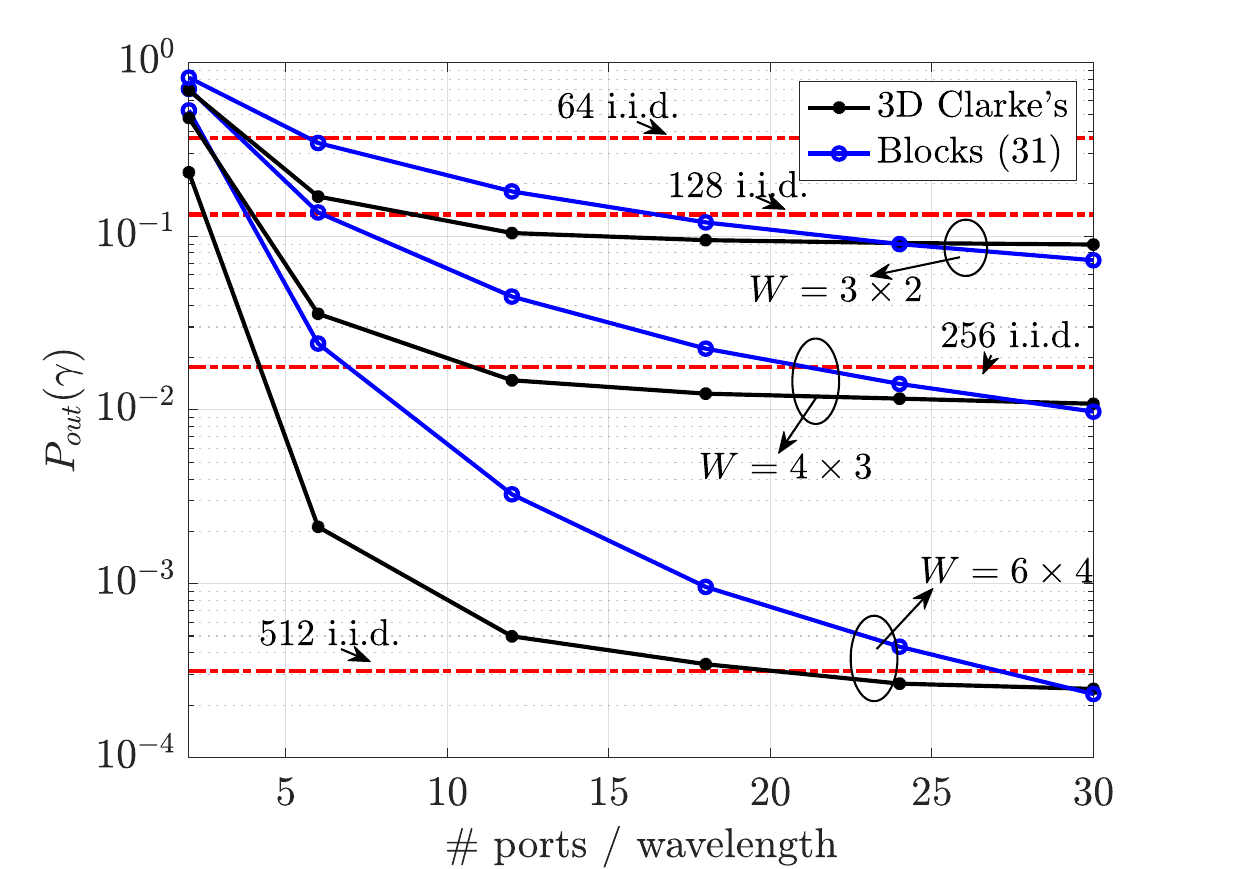}
    \caption{Evolution of outage probability as the 2D fluid aperture is densified; 3D Clarke's model \eqref{eq:3DClarke} is compared with the proposed approximation; $U = 7$, $\gamma = 1$, $\mu^2 = 0.96, \rho_\text{th} = 1$.}
    \label{fig:Nevo_2D_W}
\end{figure}

As in Fig. \ref{fig:Nevo_1D_Clarke}, the saturation effect and the performance comparison with independent antenna systems are now evaluated for 2D \gls{FAS} in Fig. \ref{fig:Nevo_2D_W}, where similar conclusions are drawn. First, we observe how the system gain saturates after a certain number of ports per wavelength, albeit the 2D aperture seems to benefit from a denser mesh of ports as compared with linear apertures; recall that in the 1D case, 10 to 15 ports per wavelength was enough to notice the saturation effect. Second, the performance gap w.r.t. the independent antenna system is larger as the fluid antenna size increases, in agreement with the analysis of Section \ref{subsec:Outage}. Last, we also see that the approximation in \eqref{eq:Pout_Quadrature} gets tighter as the antenna is densified.

\begin{figure}[t]
    \centering
    % Trim option due to figure saved directly from Matlab
    \includegraphics[trim = {0 7.5cm 0 7.5cm}, clip, width = \columnwidth]{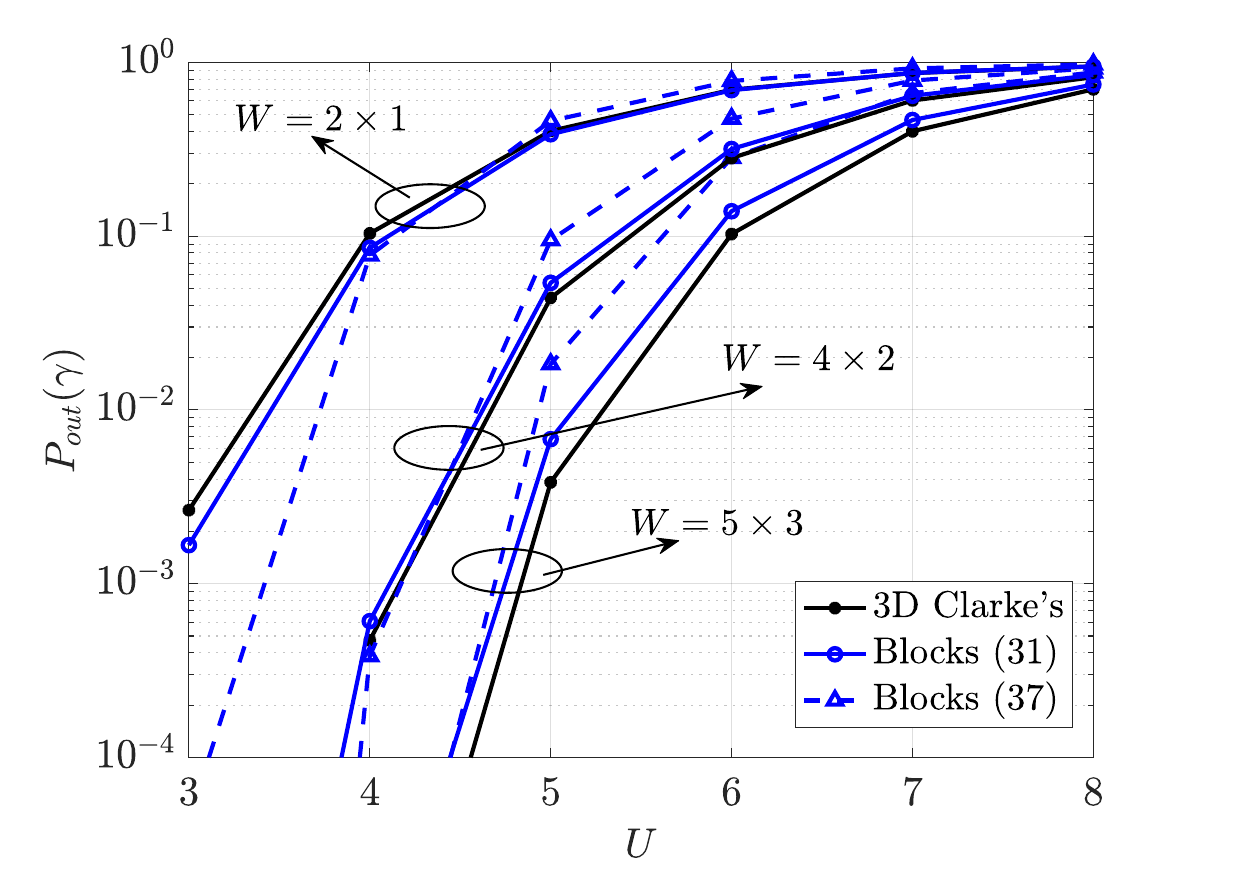}
    \caption{Evolution of the outage probability as the number of users increases; 3D Clarke's model \eqref{eq:3DClarke} is compared with the proposed approximation; $20$ ports per wavelength, $\gamma = 2$, $\mu^2 = 0.95, \rho_\text{th} = 1$.}
    \label{fig:Uevo_2D_W}
\end{figure}

Finally, Fig. \ref{fig:Uevo_2D_W} evaluates the \gls{OP} of slow-\gls{FAMA} as the number of users increases, showing a quick saturation of the system due to the added interference. Naturally, larger fluid antennas accommodate a larger number of users, but the performance rapidly drops as more users are added. Nonetheless, reasonably robust multiple access can still be provided for quite a few users (5 or 6 with $W=5\times 3$). It is important to recall that no precoding nor interference cancellation techniques are used, so the multiplexing capacity is still very relevant. We also observe that the block approximation in \eqref{eq:Pout} is once again tight regardless of the system parameters, although the simplified expression in \eqref{eq:Pout_simplified} deviates from the ground-truth performance for some particular settings. 

\begin{figure}[t]
    \centering
    % Trim option due to figure saved directly from Matlab
    \includegraphics[trim = {0 7.5cm 0 7.5cm}, clip, width = \columnwidth]{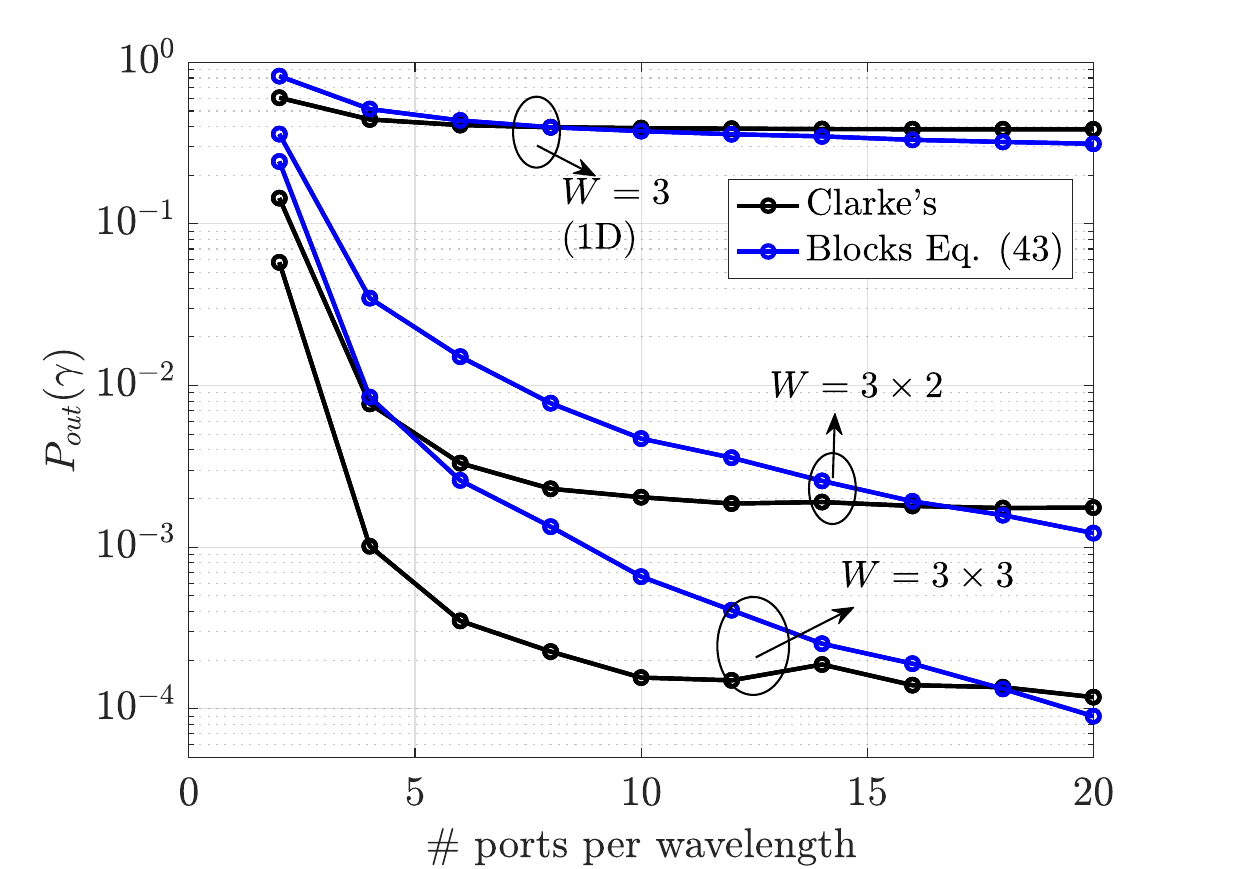}
    \caption{\rev{Evolution of OP as the fluid antenna is densified for a single user. 3D Clarke's model \eqref{eq:3DClarke} is compared with the proposed approximation; $\gamma = 5$, $\mu^2 = 0.95,$ for the 1D case and $\mu^2 = 0.98$ otherwise, $\rho_\text{th} = 1$.}}
    \label{fig:Nevo_1user}
\end{figure}

\rev{\subsection{Single-user FAS}
As a final analysis, we evaluate in Fig. \ref{fig:Nevo_1user} the saturation of \gls{FAS} in the case of a single user for different apertures (including 1D and 2D fluid antennas). As previously, we see how the proposed block-diagonal approximation gets tighter as the aperture is densified. Further, if we compare this result with Figs. \ref{fig:Nevo_1D_Clarke} and \ref{fig:Nevo_2D_W}, we can notice how the system saturates faster, i.e., less ports per wavelength are needed to reach the full performance. This is an interest result, that may suggest that increasing the number of ports beyond a certain limit is mainly beneficial to minimize interference (increases the probability of finding minima) rather than to maximize the desired channel gain. }

%------------------------------------------------------------------------------------------------------------------
% CONCLUSIONS
%------------------------------------------------------------------------------------------------------------------
\section{Conclusions}
\label{sec:Conclusions}

The proposed spatial block-correlation model has been showcased as a relevant workaround to alleviate the prohibitive analytical complexity of realistic correlation models such as Clarke's or Jakes's. Inspired by the widely adopted (temporal) block fading assumption and supported by asymptotic results on large Toeplitz matrices, the modeling framework has proved able to tightly approximate the performance of slow-\gls{FAMA} systems without penalizing the tractability of other state-of-the-art (oversimplified) models which fail in predicting realistic results. The block-correlation approximation is versatile to fit any arbitrary spatial correlation structure of the wireless channel, and is generally applicable to 1D or 2D \gls{FAS}. It enables the theoretical analysis, yielding fundamental insights on the performance limits, and greatly facilitates the numerical assessment (simulation) of \gls{FAS}. Indeed, assessing the performance via simulations under classical models such as Clarke's is highly complex, particularly for very large numbers of antenna ports (more so in 2D \gls{FAS}).      
%the use of block-diagonal correlation matrices to characterize \gls{FAS} is well supported by asymptotic results on large Toeplitz and can approximate the performance of slow-\gls{FAMA} systems tightly without penalizing the tractability of other state-of-the-art (oversimplified) models that fail in predicting realistic results.
%Although the hyperparameters of the model (namely, correlation within each have been chosen somewhat heuristically, this paper proves the potential of the block-correlation approximation and its versatility to fit any arbitrary correlation structure of the wireless channel. 
The proposed framework thus arises as a tool to fully answer fundamental questions and to understand the limits of \gls{FAS}, such as the saturation of the gain brought by apertures with denser resolution. Indeed, understanding the performance limits of \gls{FAS} is still necessary before facing other challenges such as its integration with other technologies and practical physical implementations, and the proposed modeling framework shows great potential as a means to that end.

%The fact that a completely different correlation structure successfully captures the performance under realistic (physics-based) correlation models also points out that further studies are necessary to fully understand the impact of spatial correlation in \gls{FAS}. Understanding the performance limits of \gls{FAS} is still necessary before facing other challenges such as its integration with other technologies and practical physical implementations, and given the high complexity inherent to spatial correlation models, alternative tools such as the proposed framework may become feasible solutions.
%and interesting pathways to explore.

%------------------------------------------------------------------------------------------------------------------
% Appendices
%------------------------------------------------------------------------------------------------------------------
\appendices

\section{Eigenvalues of block-diagonal matrices}
\label{app:EigBlockMatrix}
%Given the matrix structure in \eqref{eq:SigmaHat}, 
The eigenvalues of $\widehat{\bm{\Sigma}}$ are those of the block matrices $\mat{A}_1,\dots, \mat{A}_B$. For each block, these are given as solutions to
%The eigenvalues of each block $\mat{A}_b$ are obtained as the solutions to
\begin{equation}
    \left|\mat{A}_b - \rho\mat{I}\right| = \left|\begin{smallmatrix}
    1-\rho & \mu_b^2 & \cdots & \mu_b^2 \\
    \mu_b^2 & 1-\rho & \cdots & \mu_b^2 \\
    \vdots & & \ddots & \vdots \\
    \mu_b^2 & \cdots & \mu_b^2 & 1-\rho
    \end{smallmatrix}\right| = 0.
\end{equation}
%Since the determinant is invariant to linear rows/columns combinations, we successively subtract to each row the one immediately below, leading to  
%\begin{equation}
%    \left|\left(\begin{smallmatrix}
%    1-\rho - \mu_b^2 & \mu_b^2 +\rho - 1 & 0 &\cdots & 0 \\
%    0 & 1-\rho -\mu_b^2 & \mu_b^2 +\rho - 1 & \cdots & 0 \\
%    \vdots &  & & \ddots & \vdots \\
%    \mu_b^2 &  \mu_b^2 & \cdots &\mu_b^2 & 1-\rho
%    \end{smallmatrix}\right)\right| = 0.
%\end{equation}
%Similarly, we now add to each column the one immediately to the left, yielding
Since the determinant is invariant to linear rows/columns combinations, we successively subtract to each row the one immediately below, and then successively add to each column the one immediately to the left, leading eventually to 
\begin{equation}
    \left|\left(\begin{smallmatrix}
    1-\rho - \mu_b^2 & 0 & 0 &\cdots & 0 \\
    0 & 1-\rho -\mu_b^2 & 0 & \cdots & 0 \\
    \vdots &  & & \ddots & \vdots \\
    \mu_b^2 & 2\mu_b^2 & \cdots & (L_b-1)\mu_b^2 & 1-\rho + (L_b-1)\mu_b^2
    \end{smallmatrix}\right)\right| = 0.
\end{equation}
Block $\mat{A}_b$ thus gives a multiple eigenvalue $\rho = 1-\mu_b^2$ (multiplicity $L_b-1$) and a simple eigenvalue $\rho = (L_b-1)\mu_b^2 + 1$.

\section{Proof of Lemma \ref{lemma:Pout}}
\label{app:Outage}
The outage probability in \eqref{eq:Pout} is obtained by following similar steps as in \cite[Appendix B]{Wong2022} for each block of the approximated correlation matrix. The superscripts ${( \widetilde{u} ,u)}$ and $(u,u)$ are omitted for ease of notation.

\subsection{Joint PDF and CDF of $\{X_n\}$}

Conditioned on $r_b \triangleq x_b^2 + y_b^2$ $\forall$ $b$, $X_n$ is a non-central Chi-square random variable with two degrees of freedom---equivalently, a squared Rician random variable---with \gls{PDF} 
\begin{equation}
	f_{X_n | r_b}(x) = \frac{1}{2}\exp{-\frac{x + \frac{\mu^2}{1-\mu^2}r_b}{2}}\besseli{\sqrt{\frac{\mu^2r_bx}{1-\mu^2}}}.
\end{equation}
Since the variables in $\{X_n\}$ are all linked by $\{r_b\}$, then they are conditionally independent with \gls{PDF}
\begin{align}
	f_{\{X_n\} | \{r_b\}}(x_1,\dots,x_N) =& \prod_{b=1}^B\prod_{n\in\mathcal{N}_b}\frac{1}{2}\exp{-\frac{x_n + \frac{\mu^2}{1-\mu^2}r_b}{2}} \notag \\
    &\times\besseli{\sqrt{\frac{\mu^2r_bx_n}{1-\mu^2}}}.   \label{eq:joint_Xk_cond}
\end{align}
In \eqref{eq:joint_Xk_cond}, to clearly highlight the block correlation model, we have specifically split the product into the different blocks and the $n$ indices within each block. Hence, $\mathcal{N}_b$ denotes the set of $n$ indices corresponding to block $b$ (note that $|\mathcal{N}_b| = L_b$). To obtain the unconditioned joint \gls{PDF}, \eqref{eq:joint_Xk_cond} is averaged over the distribution of $r_b$'s, which are independent and follow exponential distributions with \gls{PDF}
\begin{equation}
	f_{r_b}(r_b) = \frac{1}{2}e^{-r_b/2},\quad b=1,\dots,B,
\end{equation}
leading to
%\begin{align}
%		f_{\{X_n\}}&(x_1,\dots,x_N)  = \int_0^\infty\cdots \int_0^\infty f_{\{X_n\} | \{r_b\}}(x_1,\dots,x_N) \notag \\
%  &\times\prod_{b=1}^Bf_{r_b}(r_b)  \dif r_1 \cdots \dif r_B \notag \\
%		&=  \prod_{b=1}^B\int_0^\infty \frac{1}{2^{L_b+1}}e^{-r_b/2}\prod_{k\in\mathcal{K}_b}\exp{-\frac{x_k + \frac{\mu_b^2}{1-\mu_b^2}r_b}{2}}\notag \\
 % &\times\besseli{\sqrt{\frac{\mu_b^2r_bx_k}{1-\mu_b^2}}} \dif r_b. \label{eq:joint_Xk}
%\end{align}
\begin{align}
		f_{\{X_n\}}&(x_1,\dots,x_N) =  \prod_{b=1}^B\int_0^\infty \frac{1}{2^{L_b+1}}e^{-r_b/2} \notag \\
  &\times\prod_{k\in\mathcal{K}_b}\exp{-\frac{x_k + \frac{\mu_b^2}{1-\mu_b^2}r_b}{2}}
  \besseli{\sqrt{\frac{\mu_b^2r_bx_k}{1-\mu_b^2}}} \dif r_b. \label{eq:joint_Xk}
\end{align}

From \eqref{eq:joint_Xk}, the joint \gls{CDF} is obtained as
\begin{align}
	&F_{\{X_n\}}(t_1,\dots,t_N) =\int_0^{t_1}\cdots \int_0^{t_N} f_{\{X_n\}}(\cdot) \dif x_1 \cdots \dif x_N \notag \\
	\overset{(a)}{=} &\prod_{b=1}^B\int_0^\infty \frac{1}{2}e^{-r_b/2}\prod_{n\in\mathcal{N}_b}\left[1-Q_1\left(\sqrt{\frac{\mu^2 r_b}{1-\mu^2}}, \sqrt{t_n}\right)\right]\dif r_b, \label{eq:JointCDF_Xk}
\end{align}
where the \gls{CDF} of the Rician distribution is identified in $(a)$.

\subsection{Joint PDF of $\{Y_n\}$}

The joint \gls{PDF} of $\{Y_n\}$ is obtained by following the same steps as in the case of $\{X_n\}$. First, conditioned on $\widetilde{r}_b \triangleq \sum_{\widetilde{u}\neq u} (x_b^{( \widetilde{u} ,u)})^2 + (y_b^{( \widetilde{u} ,u)})^2$ for $b = 1,\dots,B$, $Y_n$ is non-central $\chi^2$ with $2(U-1)$ degrees of freedom and \gls{PDF}
\begin{align}
	f_{Y_n | \widetilde{r}_b}(y) =& \frac{1}{2}\left(\frac{y(1-\mu^2)}{\mu^2\widetilde{r}_b}\right)^{\frac{U-2}{2}}\exp{-\frac{y + \frac{\mu^2}{1-\mu^2}\widetilde{r}_b}{2}}\notag \\
 &\times I_{U-2}\left(\sqrt{\frac{\mu^2\widetilde{r}_b y}{1-\mu^2}}\right).
\end{align}

Averaging over the set $\{\widetilde{r}_b\}$ (each of which is central $\chi^2$ distributed), the joint  \gls{PDF} of $\{Y_n\}$ is given by
\begin{align}
	f_{\{Y_n\}}&(y_1,\dots,y_N) = \prod_{b=1}^B\int_0^\infty \frac{\widetilde{r}_b^{U-2}e^{-\widetilde{r}_b/2}}{2^{U-1}\Gamma(U-1)} \notag \\
   &\times\prod_{n\in\mathcal{N}_b}\frac{1}{2}\left(\frac{y_n(1-\mu^2)}{\mu^2\widetilde{r}_b}\right)^{\frac{U-2}{2}} \exp{-\frac{y_n + \frac{\mu^2}{1-\mu^2}\widetilde{r}_b}{2}} \notag \\
   &\times I_{U-2}\left(\sqrt{\frac{\mu^2\widetilde{r}_b y_n}{1-\mu^2}}\right) \dif \widetilde{r}_b.  \label{eq:JointPDF_Yk}
\end{align}

\subsection{Outage probability}

From \eqref{eq:ObjFun}, the outage probability is obtained as
\begin{align}
	\widehat{P}_\text{out}(\gamma) &=  P\left(\frac{X_1}{Y_1} < \gamma,\dots, \frac{X_N}{Y_N} < \gamma,  \right)  \notag \\
	&= \int_0^\infty \cdots \int_0^\infty F_{\{X_n\} | \{Y_n\}}(\gamma y_1,\dots,\gamma y_N) \notag \\
 &\quad\quad\times f_{\{Y_n\}}(y_1,\dots,y_N) \dif y_1 \cdots \dif y_N.
\end{align}
Introducing \eqref{eq:JointCDF_Xk} and \eqref{eq:JointPDF_Yk} leads to \eqref{eq:AppOutage_Pout} at the top of next page, from which \eqref{eq:Pout} is obtained applying \cite[Corollary 1]{Wong2022}. 

\begin{figure*}[t]
\begin{align}
	\widehat{P}_\text{out}(\gamma) =& %\underbrace{\int_0^\infty \cdots \int_0^\infty}_{y_1,\cdots, y_N} \prod_{b=1}^B \int_0^\infty \int_0^\infty \frac{\widetilde{r}_b^{U-2}e^{-\frac{r_b + \widetilde{r}_b}{2}}}{2^{U}\Gamma(U-1)}\prod_{n\in\mathcal{N}_b} \frac{1}{2} \left[1-Q_1\left(\sqrt{\frac{\mu^2 r_b}{1-\mu^2}}, \sqrt{\gamma y_n}\right)\right] \left(\frac{y_n(1-\mu^2)}{\mu^2\widetilde{r}_b}\right)^{\frac{U-2}{2}} \notag \\
	%&\quad \quad \times \exp{-\frac{y_n + \frac{\mu^2}{1-\mu^2}\widetilde{r}_b}{2}} I_{U-2}\left(\sqrt{\frac{\mu^2\widetilde{r}_b y_n}{1-\mu^2}}\right) \dif r_b \dif \widetilde{r}_b \dif y_1 \cdots \dif y_N \notag \\
	\prod_{b=1}^B \int_0^\infty \int_0^\infty \frac{\widetilde{r}_b^{U-2}e^{-\frac{r_b + \widetilde{r}_b}{2}}}{2^{U}\Gamma(U-1)}\prod_{n\in\mathcal{N}_b} \left[1 -\frac{1}{2} \int_{y_n = 0}^\infty Q_1\left(\sqrt{\frac{\mu^2 r_b}{1-\mu^2}}, \sqrt{\gamma y_n}\right) \left(\frac{y_n(1-\mu^2)}{\mu^2\widetilde{r}_b}\right)^{\frac{U-2}{2}}\right. \notag \\
	&\quad \quad \times \left. \exp{-\frac{y_n + \frac{\mu^2}{1-\mu^2}\widetilde{r}_b}{2}} I_{U-2}\left(\sqrt{\frac{\mu^2\widetilde{r}_b y_n}{1-\mu^2}}\right) \right] \dif r_b \dif \widetilde{r}_b. \label{eq:AppOutage_Pout}
\end{align}
\hrulefill
\vspace*{4pt}
\end{figure*}

\section{Proof of Corollary \ref{coro:ApproxOutage}}
\label{app:ApproxOutage}
Consider first $\delta(\widetilde{r}_b)$ as an arbitrary threshold such that $[G(\gamma; r_b, \widetilde{r}_b)]^{L_b} = 1$ for ${r}_b < \delta(\widetilde{r}_b)$ and $[G(\gamma; r_b, \widetilde{r}_b)]^{L_b} = 0$ for ${r}_b > \delta(\widetilde{r}_b)$. Then, for large $L_b$, $[G(\gamma; r_b, \widetilde{r}_b)]^{L_b}$ in \eqref{eq:G_approx} can be approximated by the Heaviside step function shifted at $\delta(\widetilde{r}_b)$ and, consequently, \eqref{eq:Pout} is approximated by 
\begin{align}
    \widehat{P}_\text{out}(\gamma) \approx \prod_{b=1}^B \int_0^\infty\frac{\widetilde{r}_b^{U-2}e^{-\widetilde{r}_b/2}}{2^{U}\Gamma(U-1)}\int_0^{\delta(\widetilde{r}_b)}  e^{-r_b/2}\dif r_b \dif \widetilde{r}_b. \label{eq:appMarcum_Pout}
\end{align}
Solving the inner integral and using \cite[Eq. (3.351-3)]{Gradshteyn2007} gives \eqref{eq:Pout_simplified}.

It remains to calculate $\delta(\widetilde{r}_b)$. Since $[Q_p(\cdot)]^L$ corresponds to the product of the complementary \glspl{CDF} of $L_b$ i.i.d. non-central $\chi^2$ variables, a reasonable choice for the threshold $\delta(\widetilde{r}_b)$ can be obtained by seeking the value of ${r}_b$ for which $[G(\gamma; r_b, \widetilde{r}_b)]^{L_b} = \frac{1}{2}$. This value corresponds to the point of steepest slope, i.e., the minimum of the derivative (since $Q_p(\cdot)$ is monotonically decreasing) \cite{Guo2021}. Thus, we aim to solve
\begin{equation}
    \mathop {\text{arg min}}\limits_{z} \;\;\mathop \frac{\partial \left[Q_p (\alpha, z)\right]^L}{\partial z}.
\end{equation}

The partial derivative w.r.t. $z$ is given by
\begin{align}
    \frac{\partial \left[Q_p (\alpha, z)\right]^L}{\partial z} &= -L \left[Q_p (\alpha, z)\right]^{L-1}\frac{z^p}{\alpha^{p-1}}e^{-\frac{\alpha^2 + z^2}{2}}I_{p-1}(\alpha z) \notag \\
    &\overset{(a)}{\approx} -\frac{L}{\sqrt{2\pi}} \left[Q_p (\alpha, z)\right]^{L-1} \frac{z^{p-\frac{1}{2}}}{\alpha^{p-\frac{1}{2}}} e^{-\frac{(\alpha - z)^2}{2}}, \label{eq:app_Partial}
\end{align}
where, in $(a)$, the first term of the asymptotic expansion in \cite[Eq. (9.7.1)]{Abramowitz1964} for the Bessel's function is used. Equating the partial derivative w.r.t. $z$ of \eqref{eq:app_Partial} to zero, and applying again the first term of the expansion for $I_\nu(\cdot)$, we obtain
\begin{align}
     &-\frac{(L-1)}{\sqrt{2\pi}}\left[Q_p (\alpha, z)\right]^{L-2}\frac{z^{p-\frac{1}{2}}}{\alpha^{p-\frac{1}{2}}}e^{-\frac{(\alpha - z)^2}{2}} \notag \\
     &+ \left[Q_p (\alpha, z)\right]^{L-1}\frac{p-\frac{1}{2}}{z} + \left[Q_p (\alpha, z)\right]^{L-1}(\alpha - z) =  0.
\end{align}

For large $L$, $ \left[Q_p (\alpha, z)\right]^{L-1}\approx \left[Q_p (\alpha, z)\right]^{L-2}$ due to the sigmoid behaviour of $Q_p(\cdot)$, leading to 
\begin{align}
     -\frac{(L-1)}{\sqrt{2\pi}}\frac{z^{p-\frac{1}{2}}}{\alpha^{p-\frac{1}{2}}}e^{-\frac{(\alpha - z)^2}{2}} + \frac{p-\frac{1}{2}}{z} + \alpha - z =  0, \label{eq:app_EqMarcum}
\end{align}
where the sought value of $z$ is the solution to the above equation. To obtain an approximated closed-form solution, the first order Taylor's series can be taken around the point $z = \alpha$. This choice is justified as follows: if $z < \alpha$, then $Q_p(\alpha, z) \rightarrow 1$, while $Q_p(\alpha, z) \rightarrow 0$ if $z > \alpha$; therefore, it is reasonable to assume that the solution to \eqref{eq:app_EqMarcum} is in a relatively small neighbourhood around $\alpha$. Therefore, we get
\begin{equation}
    z \approx \alpha - \frac{\frac{-(L-1)}{\sqrt{2\pi}} + \alpha^{-1}(p-\frac{1}{2}))}{-\frac{L-1}{\sqrt{2\pi}}(p-\frac{1}{2})\alpha^{-1} - \alpha^{-2}(p-\frac{1}{2}) - 1}
\end{equation}
which, under the assumption of $\alpha \gg p$, yields 
\begin{equation}
        z \approx \alpha + \frac{p - \frac{1}{2} - \alpha\frac{L-1}{\sqrt{2\pi}}}{\frac{L-1}{\sqrt{2\pi}}(p-\frac{1}{2}) + \alpha}. \label{eq:z_th}
\end{equation}
Comparing \eqref{eq:z_th} with \eqref{eq:G_approx} leads to \eqref{eq:threshold}, completing the proof.

\section{Outage probability of i.i.d. antennas}
\label{app:OutageIID}
Considering $B$ independent antennas, in \eqref{eq:ObjFun},
\begin{align}
	X_n &= \left(x_n^{(u,u)}\right)^2	 + \left(y_n^{(u,u)}\right)^2, \\
	Y_n &= \sum_{\widetilde{u}=1, \widetilde{u} \neq u}^U\left(x_n^{( \widetilde{u} ,u)} \right)^2	 + \left(y_n^{( \widetilde{u} ,u)} \right)^2,
\end{align}
where all the involved variables are independent zero-mean Gaussian variables with unit variance. Therefore, $X_n$ is exponentially distributed with joint \gls{CDF}
\begin{equation}
    F_{\{X_n\}}(t_1,\dots, t_B) = \prod_{n=1}^B 1 - e^{-t_n/2}.
\end{equation}
Similarly, $Y_n$ is central $\chi^2$ distributed %with $2(U-1)$ degrees of freedom 
with joint \gls{PDF}
\begin{equation}
    f_{\{Y_n\}}(y_1,\dots, y_B) = \prod_{n=1}^B \frac{1}{2^{U-1}\Gamma(U-1)}y_n^{U-2}e^{-y_n/2}.
\end{equation}
The OP is then calculated as
\begin{align}
    P_\text{out}(\gamma) =& \int_0^\infty\cdots\int_0^\infty F_{\{X_n\} | \{Y_n\}}(\gamma y_1,\dots, \gamma y_B) \notag \\
    &\quad\quad\times f_{\{Y_n\}}(y_1,\dots, y_B) \dif y_1 \cdots \dif y_B,
\end{align}
which, after standard algebraic manipulations and the use of \cite[Eq. (3.351 3)]{Gradshteyn2007} yields \eqref{eq:Pout_ind}.

%------------------------------------------------------------------------------------------------------------------
% References
%------------------------------------------------------------------------------------------------------------------
%\vspace{10mm}
\bibliographystyle{IEEEtran}
\bibliography{references.bib}

\end{document}